\documentclass[pra,showpacs,floatfix]{revtex4}
\usepackage{graphicx}

\newcommand{\eps }{\varepsilon }
\begin{document}

\title{Many-body theory of positron-atom interactions}

\author{G. F. Gribakin}
\email[E-mail: ]{g.gribakin@am.qub.ac.uk}
\author{J. Ludlow}
\email[E-mail: ]{j.ludlow@am.qub.ac.uk}
\affiliation{Department of Applied Mathematics and Theoretical 
Physics, Queen's University, Belfast BT7 1NN, Northern Ireland, UK}

\date{\today }

\begin{abstract}
A many-body theory approach is developed for the problem of positron-atom
scattering and annihilation. Strong electron-positron correlations are included
non-perturbatively through the calculation of the electron-positron vertex
function. It corresponds to the sum of an infinite series of
ladder diagrams, and describes the physical effect of virtual
positronium formation. The vertex function is used to calculate the
positron-atom correlation potential and nonlocal corrections to the
electron-positron annihilation vertex. Numerically, we make use of B-spline
basis sets, which ensures rapid convergence of the sums over intermediate
states. We have also devised an extrapolation procedure that allows one to
achieve convergence with respect to the number of intermediate-state
orbital angular momenta included in the calculations. As a test, the present
formalism is applied to positron scattering and annihilation on hydrogen,
where it is exact. Our results agree with those of
accurate variational calculations. We also examine in detail
the properties of the large correlation corrections to the annihilation
vertex.
\end{abstract}


\maketitle


\section{Introduction}

The interaction of a low-energy positron with a many-electron atom is
characterised by strong correlation effects. Apart from the dynamic
polarisation of the electron cloud by the field of the positron, the
positron can also form positronium (Ps), by picking up one of
the atomic electrons. When the positron energy is below the Ps-formation
threshold, $\eps _{\rm Ps}=I+E_{1s}({\rm Ps})=I-6.8$ eV,
where $I$ is the atomic ionisation potential, positronium formation is
a virtual process. Nevertheless, its role in the positron-atom interaction
is very important \cite{Massey:64,Amusia:76,Dzuba:93,Gribakin:94}.
The main aim of this work is to develop a many-body theory approach which
accounts accurately for both correlation effects, and to test it for
positron scattering and annihilation on hydrogen.

The study of positron interaction with matter is a topic of
fundamental interest \cite{NewDir}. Positrons have also found many useful
applications. They are a very sensitive probe of the presence of defects in
materials \cite{Puska}. The recent development of a scanning positron
microscope \cite{David} may lead to an increased use of positrons for quality
control of materials, particularly in the semiconductor industry. In medicine,
positron emission tomography, or PET, has become a standard means of medical
imaging (see, e.g., \cite{oncology}). A proper understanding of how positrons
interact with matter at the fundamental level of atoms and molecules, is
therefore essential.

The interaction of low-energy positrons with atoms has presented a challenge
to the theorist for many decades. The study of positron scattering from atoms
was initially seen as a useful complement to work on electron scattering,
particularly in helping to understand the role of the exchange interaction.
However, although the exchange interaction is absent, it was quickly realised
that the positron-atom problem is more complex than the electron case.
The attractive induced polarisation potential tends to cancel and even
overcome the static repulsion of the positron by the atom at low energies. 
The positron may also temporarily capture one of the atomic electrons in a
process known as virtual positronium formation. The need to account for
these effects requires an elaborate and accurate theoretical description. 

For small systems, such as hydrogen and helium, accurate results were obtained
through the use of variational methods
\cite{Schwartz:61,Bhatia:7174,Humberston:73,Humberston:79,%
Humberston:97,VanReeth:99}.
Positron scattering from alkali atoms which have a single valence electron,
has been calculated extensively using a coupled-channel method with
pseudostates \cite{Walters}. More recently, positron and Ps interaction with
atoms with few active (valence) electrons has been studied using the
stochastic variational method (SVM) and configuration-interaction-type
approaches
\cite{Mitroy:98,Dzuba:99,Ivanov:01,Bromley:02a,Bromley:02b}. However, it is
difficult to extend these methods to larger atomic systems with many valence
electrons, e.g., the noble gases.

An attractive alternative to few-body methods is many-body theory
\cite{Fetter}. It lends itself naturally to the study of problems where an
extra particle interacts with a closed-shell target. The use of diagrams
makes this method both descriptive and intuitive, and allows one to
take many-particle correlations into account in a systematic way.
Many-body theory has been successful in the study of
photoionisation \cite{Amusia:75}, and in problems involving electrons, such as
electron-atom scattering \cite{Kelly:63,Amusia:75z,Amusia:82,Johnson:94},
negative ions \cite{Chernysheva:88,Johnson:88,Dzuba:91,Dzuba:94}, and precise
calculations of energies and transition amplitudes in heavy atoms with a
single valence electron \cite{Dzuba:87,Blundell:88}. The application of 
many-body theory to low-energy positron interactions with atoms has 
met with more difficulty.

Many-body theory utilises techniques originating in quantum field theory. It
describes the terms of the perturbation series in the interaction between
particles diagrammatically. The difficulty in applying this approach to
the interaction of positrons with atoms arises from the need to take
into account (virtual) Ps formation. Being a bound state, Ps cannot be
described by a finite number of perturbation-theory terms. Hence, an
infinite sequence of the `ladder' diagrams must be summed.

The first attempt to apply many-body theory to the positron-atom problem was
by Amusia {\it et al} in 1976 \cite{Amusia:76}, who used a crude approximate
method of accounting for virtual Ps formation in He. A better approximation for
the virtual Ps-formation contribution was devised in \cite{Gribakin:94}, and
applications to various atomic targets, including noble gases, were
reported \cite{Dzuba:95,Dzuba:96,Gribakin:96}. In particular, a reasonable
description was obtained for positron scattering from noble-gas atoms,
which highlighted the presence of positron-atom virtual levels in Ar, Kr and
Xe. On the other hand, application of the same approximation to
positron-atom annihilation showed that it was clearly deficient.

In spite of the approximate treatment of virtual Ps formation, many-body
theory calculations for Mg, Cd, Zn and Hg were the first to provide credible
evidence that positrons can bind to neutral atoms \cite{Dzuba:95}.
Two years later positron-atom binding was proved in a stochastic variational
calculation for Li \cite{Ryzhikh:97}. At present the list of atoms 
capable of binding positrons has expanded dramatically, SVM and
configuration-interaction calculations confirming positron binding to Mg,
Cd and Zn \cite{Ryzhikh:98,Mitroy:02}. This topic is now of major
interest in positron physics (see \cite{Mitroy:01,Mitroy:02} for useful
reviews).

In this paper, new techniques will be outlined that allow the exact
calculation of the electron-positron ladder diagram sequence which accounts
for virtual Ps formation. This approach enables many-body theory to provide
accurate information on the elastic scattering, annihilation and binding of 
positrons on atoms and negative ions at energies below the Ps formation
threshold.

\section{Many-body theory method}\label{method}

\subsection{Dyson equation}\label{dysonsec}

A conventional treatment of positron scattering from an $N$-electron target
would start from the Schr\"{o}dinger equation for the total wavefunction
for the $N+1$ particles. In many-body theory we start instead from the Dyson
equation (see, e.g., \cite{Fetter,Migdal}),
\begin{equation}\label{Dyson44}
(H_0 + \Sigma _\eps )\psi _\eps =\eps \psi _\eps ,
\end{equation} 
where $\psi _\eps $ is the single-particle (quasi-particle)
wavefunction of the positron, $\eps$ is its energy, and
$H_0$ is a central-field Hamiltonian of the zeroth approximation,
which describes the motion of the positron in the static field of the target.
The many-body dynamics in Eq. (\ref{Dyson44}) is represented by
$\Sigma_{\eps}$, a nonlocal energy-dependent correlation potential.
This quantity, also known as the optical potential, is equal to the 
self-energy part of the single-particle Green's function of the positron in the
presence of the atom \cite{Bell:59}. Due to its nonlocal nature
$\Sigma_{\eps}$ operates on the quasi-particle wavefunction as
an integral operator,
\begin{equation}\label{Dyson41}
\Sigma _\eps \psi _\eps =\int
\Sigma _\eps({\bf r}, {\bf r}')\psi _\eps
({\bf r}') d{\bf r}'.
\end{equation}

For hydrogen $H_0$ may simply be taken as the Hamiltonian of the
positron moving in the electrostatic field of the ground-state atom,
$H_0=-\frac{1}{2}\nabla ^2 +U(r)$, where
$U(r)=(1+r^{-1})e^{-2r}$ \cite{Landau} (we use atomic units throughout).
For systems containing more than one electron the Hartree-Fock (HF)
Hamiltonian (without exchange, for the positron) is the best choice.
The correlation potential $\Sigma _\eps $ is given by an infinite
perturbation series in powers of the residual electron-electron and
electron-positron interaction. Inclusion of the electrostatic interaction
in $H_0$ and the use of the HF approximation for the target electrons
means that the perturbation-theory expansion for $\Sigma _\eps $
starts with the 2nd-order diagrams, and that the diagrams do not contain
elements which describe the electrostatic potential \cite{note_exch}.

Owing to the spherical symmetry of the problem, Eq. (\ref{Dyson44}) can
be solved separately for each positron partial wave. So, in practice one
deals with radial quasiparticle wavefunctions, $\tilde P_{\eps l}(r)$, related
to $\psi _\eps $ by $\psi _\eps({\bf r})=r^{-1}
\tilde P_{\eps l}(r)Y_{lm}(\Omega )$, where $Y_{lm}(\Omega )$ is the spherical
harmonic for the orbital angular momentum $l$. Accordingly, the
self-energy operator is also found for each partial wave separately,
as $\Sigma _\eps ^{(l)}(r,r')$, see Eq. (\ref{eq:part}) in the Appendix.

\subsection{Correlation potential}

\begin{figure}[ht]
\includegraphics*[width=16cm]{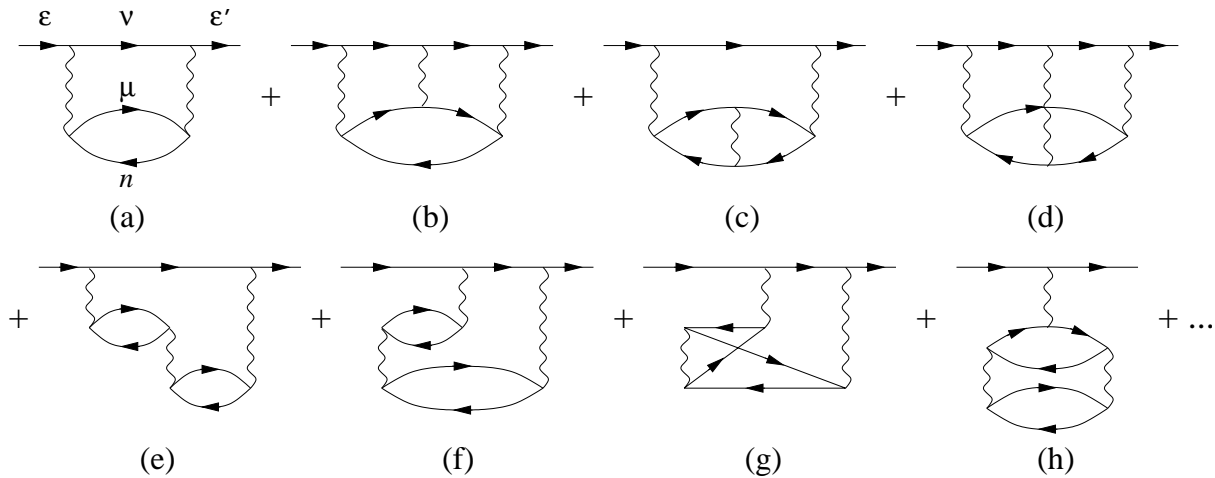}
\caption{Diagrammatic expansion of the positron-atom correlation potential
$\Sigma $. Shown are the 2nd-order and main 3rd-order contributions.
The top line in all the diagrams corresponds to the positron. Other lines
with the arrows to the right describe {\em excited} electron states, while
those with the arrows to the left correspond to holes, i.e., the electron
states {\em occupied} in the atomic ground state. Wavy lines are the
electron-positron or electron-electron Coulomb interactions. }
\label{fig:sig}
\end{figure}

Figure \ref{fig:sig} shows the lowest-order terms of the diagrammatic
expansion for the correlation potential $\Sigma $, or more precisely, for
the matrix element
$\langle \eps '|\Sigma _E| \eps \rangle $
of the correlation potential calculated at some energy $E$ between the
positron states $\eps $ and $\eps '$. The leading 2nd-order diagram, (a) in
Fig. \ref{fig:sig}, corresponds to the following expression,
\begin{equation}\label{eq:sig2}
\langle \eps '|\Sigma _E^{(2)}| \eps \rangle =
\sum _{\nu ,\,\mu ,\,n}\frac{\langle \eps 'n|V|\mu \nu \rangle
\langle \nu \mu |V|n \eps \rangle }{E-\eps _\nu -\eps _\mu +\eps _n +i0},
\end{equation}
where $V$ is the electron-positron Coulomb interaction, the sum runs over
all intermediate positron states $\nu $, excited electron states $\mu $ and
hole states $n$, and $i0$ is an infinitesimal positive imaginary quantity.

It is easy to see from Eq. (\ref{eq:sig2}) that at low energies $E$ the
diagonal matrix element $\langle\eps |\Sigma _E^{(2)}|\eps\rangle$ is negative.
This means that the 2nd-order contribution to the correlation potential,
Fig. \ref{fig:sig}(a), describes attraction.
In fact, this diagram accounts for the main correlation effect
in low-energy scattering, namely the polarisation of the atom by the
charged projectile. At large distances it leads to a well-known
local polarisation potential,
\begin{equation}\label{eq:polpot}
\Sigma _E^{(2)}({\bf r},{\bf r}')\sim
-\frac{\alpha }{2r^4}\,\delta({\bf r}-
{\bf r}') ,
\end{equation}
where $\alpha$ is the static dipole polarisability of the atom in the
HF approximation,
\begin{equation}\label{eq:alpha}
\alpha = \frac{2}{3}\sum _{\mu , n}\frac{|\langle \mu |{\bf r}|n\rangle |^2}
{\eps _\mu -\eps _n}.
\end{equation}

Besides the 2nd-order diagram, Fig. \ref{fig:sig} shows the main 3rd-order
contributions. A complete list of 3rd-order diagrams includes mirror images of
the diagrams (f) and (g). There are also a few more diagrams similar to
diagram (h), where the positron line is connected to the atomic excitation
part by a single line. They represent correlation corrections to the
HF electron charge density of the ground-state atom. Such corrections
are much smaller than other correlations effects \cite{Blundell:88},
and can be neglected.
The total number of the 3rd-order diagrams in the positron-atom
problem is considerably smaller than that in the electron case
(see, e.g., \cite{Blundell:88}), where one needs to allow for the exchange
between the incident and core electrons. 

Comparing diagrams (c), (e), (f) and (g) with (a) in Fig. \ref{fig:sig}, we
see that they represent corrections to the leading polarisation diagram,
$\Sigma ^{(2)}$, due to electron correlations within the atom.
Interaction between electron-hole pairs can in principle be included in all
orders, which would correspond to the random phase approximation (RPA)
treatment of atomic polarisation \cite{Kolb:82}. On the other hand,
if the two hole orbitals in diagrams (c) and (e) are identical, these
diagrams together with similar higher-order terms, are easily incorporated
within the 2nd-order diagram by calculating the excited electron states $\mu $
in the field of the atom with a hole in this orbital \cite{Amusia:75z}.
These approximations, and even the ``bare'' 2nd-order approximation
(with exchange diagrams added in both cases), give good results in
electron-atom scattering and negative ion problems
\cite{Amusia:82,Johnson:94,Chernysheva:88,Johnson:88,Dzuba:91,Dzuba:94}.

\begin{figure}[ht]
\includegraphics[width=14cm]{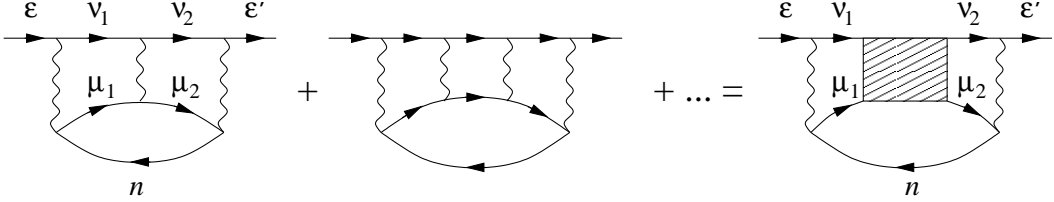}
\caption{Virtual Ps contribution to the positron-atom correlation potential
$\Sigma $.}
\label{fig:Ps}
\end{figure}

However, for the positron-atom problem the approximation based
on diagrams (a) and corrections of types (c), (e), (f), and (g), proved
to be deficient \cite{Amusia:76,Dzuba:93,Gribakin:94}. In addition one must
include the diagram Fig. \ref{fig:sig} (b) and higher-order diagrams in
which the positron interacts with the excited electron in the intermediate
state, Fig. \ref{fig:Ps}.
This sequence of diagrams accounts for virtual Ps formation. It is important
that it is summed to all orders, since in quantum mechanics a bound state
(here, Ps) which is absent in the zeroth approximation cannot be described
by a finite number of perturbation theory terms.

\begin{figure}[ht]
\includegraphics[width=16cm]{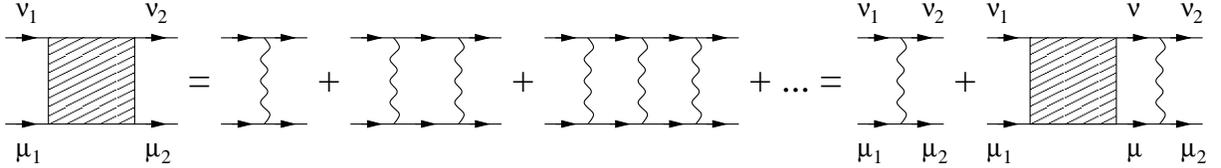}
\caption{Electron-positron ladder diagram sequence and its sum, the vertex
function $\Gamma $.}
\label{fig:lad}
\end{figure}

Summation of the diagrammatic sequence shown in Fig. \ref{fig:Ps}
is done by calculating the electron-positron {\it vertex function}
$\Gamma $, defined here as the sum of the electron-positron ladder diagrams,
Fig. \ref{fig:lad}, and denoted in the diagram by the shaded block.

Comparing the left- and right-hand sides of the
diagrammatic equation in Fig. \ref{fig:lad}, we see that the vertex function
satisfies the following linear equation,
\begin{equation}\label{eq:gamma}
\langle \nu_2\mu _2|\Gamma _E|\mu _1\nu _1\rangle =
\langle \nu_2\mu _2|V|\mu _1\nu _1\rangle + \sum _{\nu ,\,\mu }
\frac{\langle \nu_2\mu _2|V|\mu \nu \rangle 
\langle \nu \mu |\Gamma _E|\mu _1\nu _1\rangle }
{E-\eps _\nu -\eps _\mu +i0}\,.
\end{equation}
The vertex function depends on the energy $E$. It has the meaning of the
electron-positron scattering amplitude
in the field of the atom. In the lowest-order approximation $\Gamma _E=V$.

Once the vertex function is found, the contribution of virtual Ps to the
correlation potential, Fig. \ref{fig:Ps}, is obtained as
\begin{equation}\label{eq:sigG}
\langle \eps '|\Sigma _E^{(\Gamma )}| \eps \rangle =
\sum _{\nu _i,\,\mu _i,\,n}\frac{\langle \eps 'n|V|\mu _2\nu _2\rangle 
\langle \nu_2\mu _2|\Gamma _{E +\eps _n}|\mu _1\nu _1\rangle 
\langle \nu _1\mu _1|V| n \eps \rangle }
{(E-\eps _{\nu _2}-\eps _{\mu _2} +\eps _n +i0)
(E-\eps _{\nu _1}-\eps _{\mu _1} +\eps _n +i0)}\,.
\end{equation}
Since the electron-positron Coulomb interaction is attractive, $V<0$,
the terms of the perturbation series in Fig. \ref{fig:lad} have the same
sign. This explains why their sum, and hence, the contribution of
virtual Ps formation to the positron-atom attraction, are large. Note that for
electron scattering ($V>0$) this series is alternating. As a result the
net contribution of the diagrammatic series on the left-hand side in
Fig. \ref{fig:Ps} is small, and its omission in the electron-atom correlation
potential does not give rise to large errors.

The Ps-formation contribution to the correlation potential was previously
approximated by using the free $1s$-state Ps propagator orthogonalised
to the ground-state electron wavefunctions
\cite{Gribakin:94,Dzuba:96},
\begin{equation}\label{eq:Ps1s}
\langle \eps '|\Sigma _E^{(\Gamma )}| \eps \rangle \approx
\sum _n\int \frac 
{\langle \eps ' n|V |\tilde{\Psi }_{1s,{\bf K}}\rangle \langle 
\tilde{\Psi }_{1s,{\bf K}}|V|n \eps \rangle }{E+ \eps _n -
E_{1s}+K^2/4 +i0 }~\frac{d{\bf K}}{(2\pi )^3} ,
\end{equation}
where $\Psi _{1s,{\bf K}}=(8\pi )^{-1/2}\exp (-|{\bf r}-{\bf r}'|/2)
\exp [i{\bf K}\cdot ({\bf r}+{\bf r}')/2]$ is the wave
function of Ps($1s)$ with momentum ${\bf K}$, $E_{1s}
+K^2/4$ is the energy of this state, and the tilde above $\Psi _{1s,{\bf K}}$ 
in Eq. (\ref{eq:Ps1s}) indicates orthogonalisation. This approximation
is suitable for positron scattering from the targets where ground-state
Ps formation dominates, e.g., hydrogen or noble
gas atoms. It also allows one to consider positron scattering above the
Ps-formation threshold, where the correlation potential acquires an imaginary
part due to the pole in the integral in Eq. (\ref{eq:Ps1s})
\cite{Gribakin:96,Gribakin:02}.
At the same time, the ground-state Ps propagator fails to describe
the short-range electron-positron correlations crucial for the
calculation of the annihilation rates \cite{Dzuba:96}. By the uncertainty
principle, small separations invoke contributions of highly
excited states of the Ps internal motion, not included in the Ps($1s$)
propagator. In contrast, our present method based on the summation of the
ladder diagram series, Eqs. (\ref{eq:gamma}) and (\ref{eq:sigG}), is
consistent and complete. It accounts for all (virtual) intermediate
states of the electron-positron pair.

For positron scattering on hydrogen, only a few types of diagrams contribute
to $\Sigma _E$, since only one hole can be created. Moreover,
the interaction of the intermediate-state electron and positron with the
hole [diagrams (c) and (d) in Fig. \ref{fig:sig}] can be taken into
account by calculating the intermediate electron and positron wavefunctions
in the Coulomb field of the nucleus. In this case, the correlation potential
is given by the sum of the 2nd-order diagram and the virtual Ps contribution,
Figs. \ref{fig:sig}(a) and \ref{fig:Ps}, and
\begin{equation}\label{eq:sigtot}
\Sigma _E=\Sigma ^{(2)}_E+\Sigma ^{(\Gamma )}_E
\end{equation}
is the {\em exact} correlation potential. In particular, the long-range
asymptotic behaviour of $\Sigma ^{(2)}_E$ at low energies,
Eq. (\ref{eq:polpot}), contains the exact polarisability of hydrogen,
$\alpha =\frac{9}{2}$.

\subsection{Scattering}

Rather than solving the Dyson equation for the quasiparticle wavefunction
in the coordinate representation, it is easier to work with
the self-energy matrix,
\begin{equation}\label{eq:sigmat}
\langle\eps '|\Sigma _E|\eps \rangle=\int
\varphi _{\eps '}^*({\bf r}) \Sigma _E({\bf r},{\bf r}')
\varphi _\eps ({\bf r}')d{\bf r}
d{\bf r}^{\prime} ,
\end{equation}
where $\varphi _\eps $ are the positron eigenfunctions of the HF
(or ground-state hydrogen) Hamiltonian $H_0$,
\begin{equation}
H_0\varphi_{\eps}=\eps \varphi _\eps ,
\end{equation}
with a given angular momentum $l$, $\varphi _\eps ({\bf r})=r^{-1}
P_{\eps l}(r)Y_{lm}(\Omega )$. Since the static potential of the atom
is repulsive, all positron states $\varphi _\eps $ lie in the continuum
($\eps >0$). The radial wavefunctions are normalised to a $\delta $-function
of energy in Rydberg, $\delta (k^2 - {k'}^2)$, where $k$ is the positron
momentum. This corresponds to the asymptotic behaviour
\begin{equation}\label{eq:Pasym}
P_{\eps l}(r)\sim (\pi k )^{-1/2}
\sin \left( kr-l\pi /2 + \delta^{(0)}_l\right) ,
\end{equation}
where $\delta^{(0)}_l$ is the scattering phase shift in the static potential.

The matrix (\ref{eq:sigmat}) can be used to obtain the phaseshifts
directly \cite{Amusia:82}. First, a `reducible' self-energy matrix
$\langle\eps '|\tilde{\Sigma}_E|\eps \rangle$
is found via the integral equation,
\begin{equation}\label{rse}
\langle \eps '|\tilde{\Sigma}_E|\eps \rangle=
\langle\eps '|\Sigma_E|\eps \rangle + {\cal P}\int
\frac{\langle \eps' |\tilde{\Sigma}_E|\eps ''\rangle
\langle\eps ''|\Sigma_E|\eps \rangle}
{E-\eps ''}\,d\eps '',
\end{equation}
where ${\cal P}$ means the principal value of the integral. The phase shift
is then given by,
\begin{equation}\label{eq:phase}
\delta_l =\delta^{(0)}_l +\Delta \delta_l,
\end{equation}
where
\begin{equation}\label{eq:tandel}
\tan \Delta\delta_l =-2\pi \langle\eps|\tilde{\Sigma}_\eps |\eps \rangle ,
\end{equation}
determines the additional phaseshift $\Delta\delta_l(k)$ due to
correlations, at the positron energy $\eps $.

Once the reducible self-energy matrix has been found, the quasiparticle
wavefunction (also known as the {\it Dyson orbital}) can be found via,
\begin{equation}\label{eq:quasi}
\psi _\eps ({\bf r})=\varphi_ \eps ({\bf r})
+{\cal P}\int \varphi_{\eps '}({\bf r})
\frac{\langle\eps '|\tilde{\Sigma}_ \eps |\eps \rangle }
{\eps -\eps ' }d\eps ' .
\end{equation}
In order to normalise the quasiparticle radial wavefunction at large
distances to
\begin{equation}\label{nor}
\tilde P_{\eps l}(r)\sim (\pi k )^{-1/2}
\sin \left( kr- l\pi /2 + \delta^{(0)}_l+ \Delta\delta_l \right) ,
\end{equation}
the function obtained from the right-hand side of Eq.~(\ref{eq:quasi})
must be multiplied by the factor,
\begin{equation}\label{nor1}
\cos\Delta\delta_l=\left[ 1+
\left(2\pi\langle\eps|\tilde{\Sigma}_{\eps}|\eps\rangle\right)^2\right]^{-1/2}.
\end{equation}

\subsection{Positron annihilation}\label{subsec:ann}

The annihilation rate $\lambda $, of a positron in a gas of number density $n$
is usually expressed in terms of the effective number of electrons,
$Z_{\rm eff}$, which contribute to annihilation on an atom or molecule
\cite{Fraser:68},
\begin{equation}\label{Zeff}
{\lambda}={\pi} r_0^2cnZ_{\rm eff} ,
\end{equation}
where $r_0$ is the classical electron radius and $c$ is the speed of
light. Equation (\ref{Zeff}) defines $Z_{\rm eff}$ as the ratio of the
positron two-photon annihilation cross section of the atom to the
spin-averaged two-photon annihilation cross section of a free electron
in the Born approximation \cite{QED}. Annihilation takes place at
very small electron-positron separations, $\hbar /(mc)\sim 10^{-2}$~a.u.
Hence, for nonrelativistic positrons it is determined by the electron
density at the positron, and $Z_{\rm eff}$ can be calculated
as \cite{Fraser:68},
\begin{equation}\label{Zeff1}
Z_{\rm eff}=\sum_{i=1}^N\int\left|\Psi ({\bf r}
_1,{\bf r}_2,\dots ,{\bf r}_N,{\bf r})\right|^2
{\delta}({\bf r}_i-{\bf r})d{\bf r}_1
\dots d{\bf r}_N d{\bf r},
\end{equation}
where $\Psi ({\bf r}_1,{\bf r}_2,\dots ,{\bf r}_N,{\bf r})$ is the full
$(N+1)$-particle wavefunction of the $N$ electron coordinates ${\bf r}_i$ 
and positron coordinate ${\bf r}$. The wavefunction is normalised to a 
positron plane wave at large positron-atom separations,
\begin{equation}\label{eq:large}
\Psi ({\bf r}_1,{\bf r}_2,\dots ,{\bf r}_N,{\bf r})\simeq
\Phi _0({\bf r}_1,{\bf r}_2,\dots ,{\bf r}_N)e^{i{\bf k}\cdot {\bf r}},
\end{equation}
where $\Phi _0({\bf r}_1,{\bf r}_2,\dots ,{\bf r}_N)$ is the atomic
ground-state wavefunction, and ${\bf k}$ is the incident positron
momentum.

\begin{figure}[ht]
\includegraphics*[width=16cm]{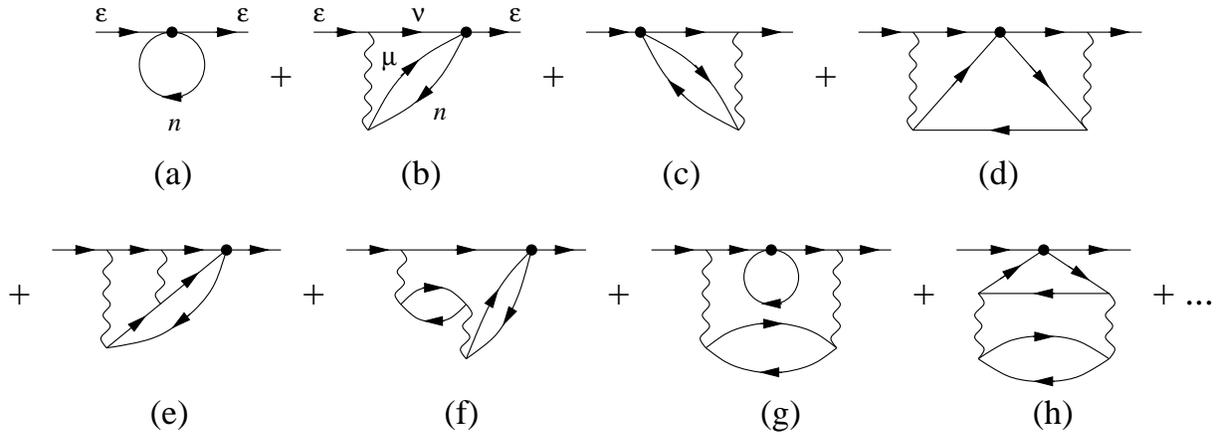}
\caption{Diagrammatic expansion of $Z_{\rm eff}$. Apart from the 0th and 1st
order diagrams (a), (b) and (c), the main types of 2nd-order diagrams are
shown. The external lines of these diagrams ($\eps$) represent the
wavefunction of the incident positron. The solid circle corresponds to the
electron-positron $\delta $-function annihilation vertex.}
\label{fig:Zeff}
\end{figure}

Although $Z_{\rm eff}$ is basically a cross section, Eq. (\ref{Zeff1}) has
the form of a transition amplitude. This enables one to apply the apparatus of
many-body theory to this quantity directly \cite{Dzuba:93,Dzuba:96}.
In this ``transition amplitude'' the initial and final states are identical,
and the electron-positron two-body operator,
$\sum _i\delta ({\bf r}_i-{\bf r})$, plays the role of a perturbation.
The positron energy in the initial and final states is the same,
$\eps =k^2/2$, and (owing to the spherical symmetry of the target) the
perturbation conserves the positron angular momentum $l$.
Therefore, the contribution of each positron partial wave to
$Z_{\rm eff}$ can be determined separately. The corresponding many-body
diagrammatic expansion is presented in Fig. \ref{fig:Zeff}.

The analytical expression for the 0th-order diagram, Fig. \ref{fig:Zeff}(a),
is
\begin{equation}\label{eq:Zeff0}
Z_{\rm eff}^{(0)}=\sum _n \int \psi _\eps ^*({\bf r})\varphi _n^*({\bf r}_1)
\delta ({\bf r}-{\bf r}_1)\varphi _n({\bf r}_1)\psi _\eps ({\bf r})
d{\bf r}d{\bf r}_1=
\sum _n \int |\varphi _n({\bf r})|^2|\psi _\eps ({\bf r})|^2 d{\bf r},
\end{equation}
where $\psi _\eps $ is the positron wavefunction, $\varphi _n$ is the
wavefunction of the hole, and the sum over $n$ runs over all holes, i.e.
orbitals occupied in the target ground state. This contribution is simply an
overlap of the electron and positron densities,
$\sum _n|\varphi _n({\bf r})|^2$ and $|\psi _\eps ({\bf r})|^2$, respectively.

The two 1st-order ``corrections'', Fig. \ref{fig:Zeff}(b) and (c), are
identical, and their contribution is
\begin{equation}\label{eq:Zeff1}
Z_{\rm eff}^{(1)}=2 \sum _{\nu ,\,\mu ,\,n}\frac{\langle \eps n|\delta |
\mu \nu \rangle \langle \nu \mu |V|n \eps \rangle }
{\eps -\eps _\nu -\eps _\mu +\eps _n },
\end{equation}
cf. Eq. (\ref{eq:sig2}) for the 2nd-order contribution to the correlation
potential. In the calculations of annihilation we assume that
the positron energy is below all other inelastic thresholds, hence, we
have dropped $i0$ in the energy denominator. Physically, 
the 1st-order diagram describes positron annihilation with an electron
``pulled out'' from the atom by the positron's Coulomb field. Calculations
in Refs. \cite{Dzuba:93,Dzuba:96} showed that for noble-gas atoms
the size of the 1st-order corrections is approximately equal to the
0th-order contribution. This means that higher-order terms must also be taken
into account.

Diagrams (d)--(h) in Fig. \ref{fig:Zeff} illustrate the main types of 2nd-order
corrections to the {\em annihilation vertex}. It is also important
to consider corrections to the incident positron wavefunctions denoted by
$\eps $. However, these corrections are included in all orders in the positron
quasiparticle wavefunction, obtained from the Dyson equation (\ref{Dyson44}),
or via equation (\ref{eq:quasi}). Hence their contribution
to $Z_{\rm eff}$ is accounted for by using the positron Dyson orbitals
$\psi _\eps $ in the calculation of the annihilation diagrams.

\begin{figure}[ht]
\includegraphics*[width=16cm]{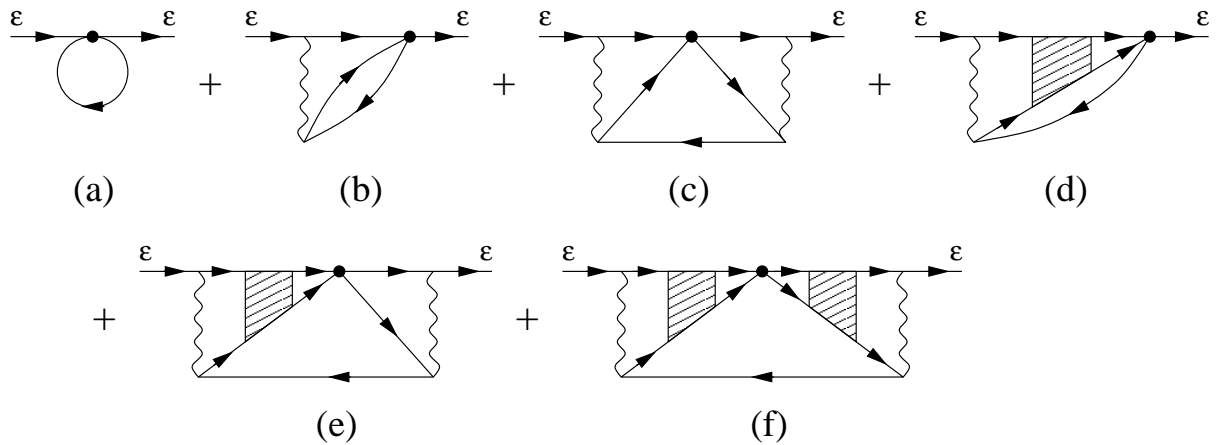}
\caption{Main contributions to the positron-atom annihilation parameter
$Z_{\rm eff}$. To account for the mirror images of the diagrams (b), (d) and
(e), their contributions are multiplied by 2.}
\label{fig:Zeff1}
\end{figure}

The presence of the $\delta $-function operator in the annihilation diagrams
enhances the importance of small electron-positron separations. For this
reason, the most important diagrams in $Z_{\rm eff}$ are those with the
Coulomb interactions between the annihilating pair, e.g., the 2nd-order
diagrams (d) and (e) in Fig. \ref{fig:Zeff} (the latter together
with its mirror image), and similar higher-order terms. A complete
all-order calculation of their contribution is achieved by using the vertex
function, as shown in Fig. \ref{fig:Zeff1}. Note that for hydrogen this
set of diagrams is exhaustive, provided the intermediate electron and positron
states are calculated in the field of the bare nucleus. Previously
diagrams (c)--(f) in Fig. \ref{fig:Zeff1} have only been
estimated \cite{Dzuba:96}. We will see that the ability to calculate these
diagrams accurately is crucial for obtaining correct values of $Z_{\rm eff}$.
We will also see in Sec. \ref{sec:res} that the role of the vertex function
(representing virtual Ps) in annihilation is much greater than in
scattering.

Of course, for many-electron targets one can also consider other diagrams,
e.g., (f)--(h) in Fig. \ref{fig:Zeff}. In particular, diagram (f)  describes
screening of the electron-positron Coulomb interaction by other electrons
[cf. Fig. \ref{fig:sig} (e)]. Diagram (g) can be viewed as the lowest-order
``pick-off'' annihilation contribution. Here the positron excites an
electron-hole pair (a precursor of virtual Ps formation) and annihilates with
an electron from one of the ground-state orbitals. Diagram (h) is independent
of the positron energy. It represents one of the corrections to the
HF ground-state electron density, cf. Fig. \ref{fig:sig}(h).
Unlike the diagrams in Fig. \ref{fig:Zeff1}, these contributions are
not systematically enhanced by the electron-positron Coulomb interaction
at small distances.

It is clear from Fig. \ref{fig:Zeff} that most
correlation corrections to the annihilation vertex, including the dominant
sequence of diagrams in Fig. \ref{fig:Zeff1}, are {\em nonlocal}.
As a result, the total $Z_{\rm eff}$ can be written as
\begin{equation}\label{eq:nonloc}
Z_{\rm eff} = \int \sum_n |\varphi_n({\bf r})|^2|\psi_{\eps}({\bf r})|^2
d{\bf r} +\int \psi_\eps ^*({\bf r})
\Delta _\eps ({\bf r},{\bf r}')\psi_ \eps ({\bf r}') d{\bf r}d{\bf r}' ,
\end{equation}
where $\Delta _\eps ({\bf r},{\bf r}')$ represents the nonlocal
correlation correction to the annihilation vertex. In the
approximation of Fig. \ref{fig:Zeff1}, it is equal to the sum of all
diagrams (b)--(f) with the external positron lines $\eps $ detached.
Diagram (h) in Fig. \ref{fig:Zeff} and similar diagrams which represent
corrections to the HF electron density, could be included by
replacing of the HF electron density $\sum_n |\varphi_n({\bf r})|^2$
in Eq. (\ref{eq:nonloc}) with the exact target electron density
$\rho _e({\bf r})$. Given the high accuracy of the
HF density, this would make only a small change in $Z_{\rm eff}$.
The structure of Eq. (\ref{eq:nonloc}) shows that even when one uses the
best single-particle positron wavefunction, the annihilation rate is not
reduced to a simple (local) overlap of the electron and positron densities.

There is an important physical difference between the correlation effects
in positron scattering and annihilation.
The key role played by the long-range polarisation potential for low-energy
positrons means that large distances are important. Polarisation also
emphasises the role played by the dipole part of the positron-target Coulomb
interaction and dipole target excitations. The contribution of
virtual Ps formation to $\Sigma _E$ is a short-range effect. The typical
distances here are comparable to the radius of the atom, or the radius of
ground-state Ps. The net effect of the strong positron-atom attraction
brings about low-lying virtual $s$-states (see, e.g., \cite{Landau}) for
Ar, Kr and Xe \cite{Dzuba:96}, or positron-atom weakly bound states,
e.g., in Mg \cite{Dzuba:95,Gribakin:96}. In both cases, the positron
scattering phaseshifts and the positron quasiparticle wavefunction in the
vicinity of the atom vary rapidly as functions of the positron energy.

On the contrary, in the annihilation diagrams the $\delta $-function vertex
emphasises small electron-positron separations. By the uncertainty principle,
such small separations correspond to high-energy excitations in the
intermediate states in the diagrams (b)--(f), Fig. \ref{fig:Zeff1}.
As a result, the non-local correction to the
annihilation vertex, $\Delta _\eps$, has a weak energy dependence, and the
energy dependence of the 2nd term in Eq. (\ref{eq:nonloc})
is almost entirely due to that of the positron wavefunctions.
The only exception is when the positron energy approaches the
Ps-formation threshold from below. Here the virtual-Ps contribution,
Fig. \ref{fig:Zeff1}(f), rises sharply. Details of the threshold behaviour
of $Z_{\rm eff}$ are discussed in Ref. \cite{threshold}.

The large difference in the energy scales characteristic of positron
scattering and annihilation has another physically important consequence.
It turns out that the {\em relative} size of the annihilation vertex
corrections in $Z_{\rm eff}$, i.e. the ratio of the 2nd term in
Eq. (\ref{eq:nonloc}) to $Z_{\rm eff}^{(0)}$, is about the same, whether
$\psi _\eps $ are the positron wavefunctions in the repulsive static
potential, $\psi _\eps =\varphi _\eps $, or the Dyson orbitals which fully
account for the positron-atom correlation potential. Numerical illustrations
of this effect will be provided in Sec. \ref{sec:res}.

Finally, we should mention that the correct normalisation of $Z_{\rm eff}$
for positrons with angular momentum $l$ is obtained by multiplying the
diagrams in Figs. \ref{fig:Zeff} and \ref{fig:Zeff1} by the extra numerical
factor,
\begin{equation}\label{eq:fact}
\frac{4\pi^2}{k}(2l+1) .
\end{equation}
This follows from the structure of the positron wavefunction
$\psi _{\bf k}$, which has the asymptotic behaviour of a plane wave
$e^{i{\bf k}\cdot {\bf r}}$ at large distances [cf. Eq. (\ref{eq:large})],
\begin{equation}\label{eq:psi_k}
\psi _{\bf k}=\frac{4\pi}{r}\sqrt{\frac{\pi }{k}}\sum _{l=0}^\infty
\sum _{m=-l}^{m=l}i^le^{i\delta _l}\tilde P_{\eps l}(r)
Y_{lm}^*(\Omega _{\bf k}) Y_{lm}(\Omega _{\bf r}) .
\end{equation}
To derive Eq. (\ref{eq:fact}), one can use $\psi _{\bf k}$ as the
external positron lines in an annihilation diagram, and perform averaging over
the directions of ${\bf k}$.


\section{Numerical implementation}

\subsection{Use of B-splines and convergence}

To evaluate the diagrams of the correlation potential $\Sigma $ and
annihilation parameter $Z_{\rm eff}$, one first needs to generate sets of
electron and positron HF basis states. These are then used to
calculate matrix elements of the Coulomb and $\delta $-function operators,
the main building blocks of the diagrams. Evaluation of the diagrams
requires summation over complete sets of electron and
positron intermediate states, including integration over
the electron and positron continua.

To perform a numerical calculation, the continuous spectrum can be discretised.
The simplest way of doing this is by placing the system in a spherical
cavity of radius $R$. Setting the wavefunctions to zero at the boundary
will result in a discrete spectrum of eigenstates with an approximately
constant stepsize in momentum space,
\begin{equation}\label{eq:quant}
\Delta k\approx \pi /R.
\end{equation}
If the value of $R$ is sufficiently large ($R\gg R_{\rm at}$, where
$R_{\rm at}\sim  1$~a.u. is the size of the  atom), the
presence of the boundary will not affect the quantities calculated. Indeed,
for the positron energy below all inelastic thresholds (except, of course,
annihilation), the intermediate states in the diagrams are {\it virtual}, and
no particle in an intermediate state can escape to infinity.

The drawback of this procedure is that for a suitably large $R$, the stepsize
in momentum is small, e.g., for $R=30$~a.u., $\Delta k\approx 0.1$~a.u. Hence,
one would need large numbers of intermediate states
to achieve convergence. Note that the actual upper energy limit depends on
the quantity in question. Thus, diagrams in $Z_{\rm eff}$ converge more slowly
than those of the correlation potential $\Sigma $, because of the greater
role of small electron-positron separations and high orbital angular momenta
in the former. However, as a rough guide, summing up to the energy of
$10^2$~a.u. should be sufficient for both. The question of the number of
intermediate states is
especially important for the calculation of the vertex function $\Gamma $,
which is a $N_{\Gamma}\times N_{\Gamma}$ matrix [see Eq. (\ref{eq:gamma})],
where $N _{\Gamma}\sim N^2(l_{\rm max}+1)$, $N$ being the number of electron
or positron states in each partial wave and $l_{\rm max}$ being the largest
orbital angular momentum included. It is clear that here the simple cavity
quantisation cannot work.

Instead, to achieve an accurate and economical span of the continuum we use
B-splines \cite{Sapirstein:96}. B-splines of order $k$ are $n$ piecewize
polynomials of degree $k-1$ defined by a knot sequence $r_j$ which divides
the interval $[0,R]$ into $n-k+1$ segments \cite{deBoor}. The basis states
are obtained by expanding the radial wavefunctions $P_l(r)$ in terms of
B-splines $B_i(r)$,
\begin{equation}\label{eq:Bspline}
P_l(r)=\sum C_i^{(l)}B_i(r),
\end{equation}
and finding the eigenvectors and eigenvalues of the radial part of the
HF (or hydrogen atom) Hamiltonian for each orbital angular
momentum $l$ by solving the generalised eigenvalue problem,
\begin{equation}\label{eq:geneig}
\sum H_{ij}C_j^{(l)}=\eps \sum Q_{ij}C_j^{(l)},
\end{equation}
where $H_{ij}=\langle B_i|H_0^{(l)}|B_j\rangle $, and
$Q_{ij}=\langle B_i|B_j\rangle $. Prior to solving Eq. (\ref{eq:geneig}), the
ground-state atom HF Hamiltonian is generated by a conventional
HF routine \cite{ATOM}. Note that in the sums over $i$ and $j$
in Eqs. (\ref{eq:Bspline}) and (\ref{eq:geneig}) the first and last splines
are discarded to implement the boundary condition $P_l(0)=P_l(R)=0$, leaving
one with a set of $n-2$ eigenstates for each electron and positron orbital
angular momentum. When Eq. (\ref{eq:geneig}) is solved for the
electron, it yields the wavefunctions of the orbitals occupied in the atomic
ground state (holes), as well as those of the excited states
(particles). The exact energies of the excited electron
and positron states are determined by the B-spline radial knot
sequence.

It is instructive to try to design an ideal distribution of energies
of a discrete set spanning the continuum. Qualitatively, at low
energies ($\eps \ll 1$~a.u.) the continuous spectrum states oscillate slowly,
and the contribution of large distances in the matrix elements is important
(hence, the need for a large $R$). As the energy of the states increases,
the range of important distances becomes smaller and smaller. Indeed, the
matrix elements then contain rapidly oscillating factors of $e^{ikr}$ type,
which means that the dominant contribution comes from $r\lesssim k^{-1}$.
Therefore, one does not need a large value of the cavity radius $R$ for the
higher-energy states. More specifically, one can estimate the necessary radius
as $R\sim a/k$, where $a$ is a number greater than unity.
Combining this with the cavity quantisation condition (\ref{eq:quant}),
one obtains $\Delta k/k \sim \pi /a$, which yields the following grid in
momentum space:
\begin{equation}\label{eq:kj}
k_j=k_0 e^{\beta j},
\end{equation}
where $k_0$ is the lowest momentum, $\beta =\pi /a$, and $j=0,\,1,\,2,\dots $.
Thus, it appears that the optimal momentum and energy
grids are exponential. By choosing a small initial momentum $k_0\ll 1$~a.u.
and $\beta <1$ one ensures that the stepsize in momentum,
$\Delta k\approx \beta k$, is sufficiently small, to describe accurately
the energy variation of the quantities summed.

\begin{figure}[ht]
\includegraphics[width=10cm]{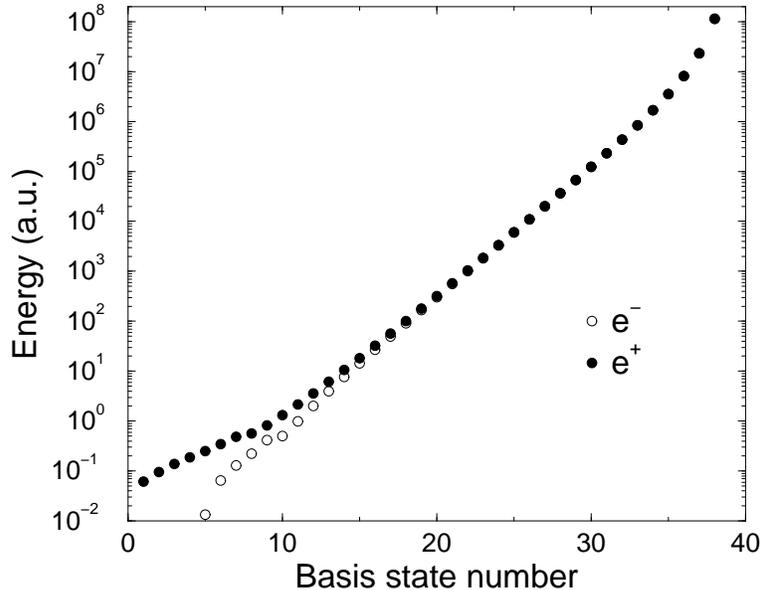}
\caption{Energies of the electron (open circles) and positron (solid circle)
$s$-wave B-spline basis states in the field of the hydrogen nucleus,
obtained using $R=30$, $k=6$ and $n=40$. The first four electron states
with negative energies ($-0.50000$, $-0.12500$, $-0.05542$ and $-0.00246$~a.u.)
are not shown.}
\label{fig:basis}
\end{figure}

It turns out that basis sets generated by Eqs. (\ref{eq:Bspline}) and
(\ref{eq:geneig}) using an exponential radial knot sequence
\cite{Sapirstein:96} are a very close realisation of the exponential
energy grid, and are {\em effectively complete}.
In the present work we use
$n=40$ B-splines of order $k=6$ with a knot sequence
\begin{equation}\label{eq:r_i}
r_j=\rho (e^{\sigma j}-1),
\end{equation}
where $\rho =10^{-3}$~a.u., and $\sigma $ is determined by the condition
$r_{n-k+1}=R$. Figure \ref{fig:basis} shows the positive
eigenenergies of the electron and positron basis states with $l=0$ for the
hydrogen atom. Their distribution does indeed correspond to the exponential
ansatz, Eq. (\ref{eq:kj}), with $\beta \approx \sigma $.
The highest energies in the sets are about $10^8$~a.u.
This value is close to the magnitude of $(\rho \sigma )^{-2}$,
since the knot point closest to the origin determines the most rapidly
varying eigenstate (by the uncertainty principle). In fact,
it may not be necessary to include all 38 basis states in each partial wave in
the many-body theory sums. In the calculations reported in this paper we
use only about 15 lowest states, which span the energy range from threshold to
$\sim 10^2$~a.u. 

Note that B-spline basis sets are used widely in atomic
physics \cite{Sapirstein:96}, and that there are other basis sets which
show a near-exponential spanning of the continuum. In particular, Laguerre
basis states provide rapid convergence in close-coupling electron-atom
scattering calculations \cite{Laguerre}.


\subsection{Calculation of the self-energy and annihilation diagrams}

The self-energy and annihilation diagrams are calculated
by summation over the B-spline basis states, and the vertex function
is found by matrix inversion from Eq. (\ref{eq:gamma}).
The angular parts of the states
are separated in the matrix elements and the angular variables are integrated
over analytically. The actual expressions for the diagrams are given in the
Appendix. The self-energy matrix and the vertex function are energy dependent.
In practice, the self-energy matrix has been calculated at 8 energies spaced
evenly from zero to the Ps formation threshold. Interpolation onto any
required energy $E$ is then used. 

Apart from the B-spline basis states, we also consider true positron
continuum HF states (\ref{eq:Pasym}). They are needed to evaluate
the matrix elements $\langle \eps |\Sigma_{E}|\eps '\rangle$ and obtain the
phaseshifts via Eqs. (\ref{rse})--(\ref{eq:tandel}). Here we use 201 states
that form an equidistant mesh in positron momenta of size $\Delta k=0.02$.
A transformation of the B-spline basis matrix elements
$\langle i|\Sigma _E|j\rangle $ into $\langle \eps |\Sigma_{E}|\eps '\rangle$
could be done using the effective completeness of the B-spline states
on the interval $[0,R]$,
\begin{equation}\label{eq:complete}
\langle \eps |\Sigma_{E}|\eps '\rangle =\sum _{i\,,j}\langle \eps |i\rangle
\langle i|\Sigma _E|j\rangle \langle j|\eps '\rangle ,
\end{equation}
where $\langle \eps |i\rangle $ is the overlap of the HF state with
the B-spline basis state. However, unlike the B-spline states which satisfy
the zero boundary condition at $r=R$, the continuous spectrum state
$P_{\eps l}$ is finite at the boundary. To fix this problem
we insert a radial weighting function $f(r)=R-r$ into equation
(\ref{eq:complete}), which now reads
\begin{equation}
\label{complete1}
\langle\eps|\Sigma_{E}|\eps '\rangle=\sum_{i,\,j}\langle\eps|f|i\rangle
\langle i|f^{-1}\Sigma_{E}f^{-1}|j \rangle\langle j|f|\eps^{\prime}\rangle ,
\end{equation}
and calculate the ``weighted'' self-energy matrix $\langle i|f^{-1}
\Sigma_{E}f^{-1}|j \rangle$, rather than $\langle i|\Sigma_{E}|j \rangle$.
The singularity of $f^{-1}$ at $r=R$ does not cause a problem, since the
B-spline basis states in the Coulomb matrix elements
involved, vanish at $r=R$. The same trick is applied in the calculation of the
annihilation diagrams.

To calculate $\langle\eps|\Sigma_{E}|\eps '\rangle $ more accurately
at low positron energies, where distances beyond $r=R$ can be important, we
make use of the long-range asymptotic form of the correlation
potential (\ref{eq:polpot}). The contribution of $r>R$ can be evaluated
as
\begin{equation}\label{eq:larger}
\int _R ^\infty P_{\eps l}(r)\left( -\frac{\alpha }{2r^4}\right)
P_{\eps ' l}(r)dr,
\end{equation}
with the correct value of the dipole polarisability $\alpha $,
and added to $\langle\eps|\Sigma_{E}|\eps '\rangle $.


\subsection{Convergence with respect to the orbital angular momenta}

The use of a B-spline basis means that fast convergence is achieved with
respect to the number of states with a particular orbital angular momentum.
However, this leaves open the question of convergence with respect to the
maximal orbital angular momentum of the electron and positron intermediate
states included in the calculation.
It has been known for a while that calculations of positron-atom scattering
converge slowly with respect to the number of target angular momenta
included in the expansion of the total wavefunction, notably slower than in
the electron-atom case \cite{Bray:93}. This is also true for the
configuration-interaction-type calculations of positron-atom bound states
\cite{Dzuba:99,Bromley:02a,Mitroy:99} and scattering \cite{Bromley:02b}.
Calculations of annihilation rates converge even more slowly
\cite{Bromley:02a,Bromley:02b,Mitroy:99}.
Physically, the slow convergence rate arises from the need to describe 
virtual Ps localised outside the atom by an expansion in terms
of single-particle orbitals centred on the nucleus. 

The problem of the convergence rate with respect to the maximal orbital
angular momentum has been investigated by the
authors in Ref. \cite{Ludlow}. Using a perturbation-theory approach
and the original ideas of Schwartz \cite{Schwartz:62}, we derived 
asymptotic formulae that describe the convergence of the scattering
amplitudes, or the phase shifts, and annihilation rates, or $Z_{\rm eff}$. 
The contribution of high orbital angular momenta probes small particle
separations in the system. The difference between the convergence
rates of the scattering and annihilation parameters is due to the
presence of either the Coulomb interaction or the $\delta$-function
annihilation operator in the relevant amplitudes.

The increments to the phaseshifts and $Z_{\rm eff}$ upon increasing the
maximum orbital angular momentum from $\lambda -1$ to $\lambda $,
were found to behave as $(\lambda +1/2)^{-4}$ and $(\lambda +1/2)^{-2}$,
respectively. This means that if a series of calculations is stopped at
some maximal angular momentum $\lambda =l_{\rm max}$, the values obtained
approach the ultimate ($l_{\rm max}\rightarrow \infty$) values as follows: 
\begin{equation}\label{eq:asym3}
\delta _l ^{[l_{\rm max}]}\simeq 
\delta _l-\frac{A}{(l_{\rm max}+1/2)^3} ,
\end{equation}
\begin{equation}\label{eq:asym4}
Z_{\rm eff}^{[l_{\rm max}]}\simeq
Z_{\rm eff} -\frac{B}{(l_{\rm max}+1/2)} .
\end{equation}
where $A$ and $B$ are some constants. They are determined
together with $\delta _l$ and $Z_{\rm eff}$ by fitting Eqs. (\ref{eq:asym3})
and (\ref{eq:asym4}) to the numerical data obtained for a range of
$l_{\rm max}$. This extrapolation to $l_{\rm max}\rightarrow \infty$
is performed at each positron momentum value considered, and is especially
important for obtaining correct values of $Z_{\rm eff}$.


\section{Results: scattering and annihilation on hydrogen}\label{sec:res}

The theory outlined above can be readily applied to
any closed-shell atom or ion. In this paper we would like to test it for
the simplest possible target, the hydrogen atom. Since it contains only one
electron, the correlation potential from Eq. (\ref{eq:sigtot}) and the
$Z_{\rm eff}$ diagrams in Fig. \ref{fig:Zeff1} give an exact solution of the
elastic scattering and annihilation problems, provided the intermediate
electron and positron states are calculated in the field of H$^+$.
The key advance of the present many-body theory of positron-atom interactions
relates to the calculation of the electron-positron vertex
function $\Gamma $, Eq. (\ref{eq:gamma}) and its incorporation
in the self-energy and annihilation diagrams. It is mainly these features of
the theory that a positron-hydrogen calculation is intended to test.

In the numerical implementation we first generate the electron and positron
B-spline basis sets. They are then used to evaluate the matrix elements,
find the vertex function and calculate the self-energy and annihilation
diagrams (see Appendix). Using the self-energy matrix, the phaseshifts are
obtained by means of Eqs. (\ref{rse})--(\ref{eq:tandel}), and the positron
Dyson orbital is calculated from Eqs. (\ref{eq:quasi}) and (\ref{nor1}).
In the end, the Dyson orbitals replace the positron HF states
in the external lines of the annihilation diagrams, and final values of
$Z_{\rm eff}$ are obtained. To test the stability of the results with respect
to the cavity radius, the calculations were performed with $R=15$ and
30 a.u.

\begin{figure}[ht]
\includegraphics[width=8cm]{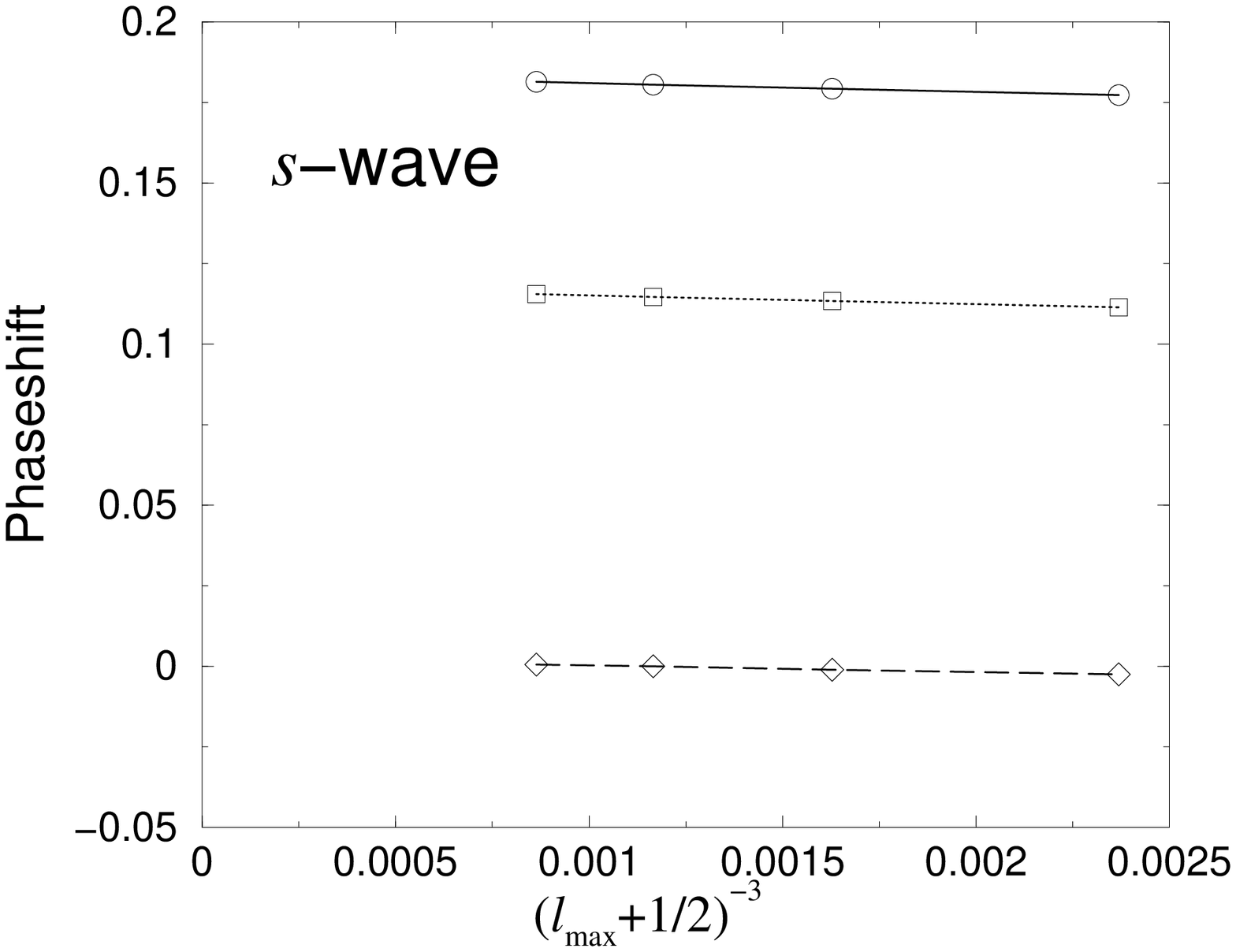}
\includegraphics[width=8cm]{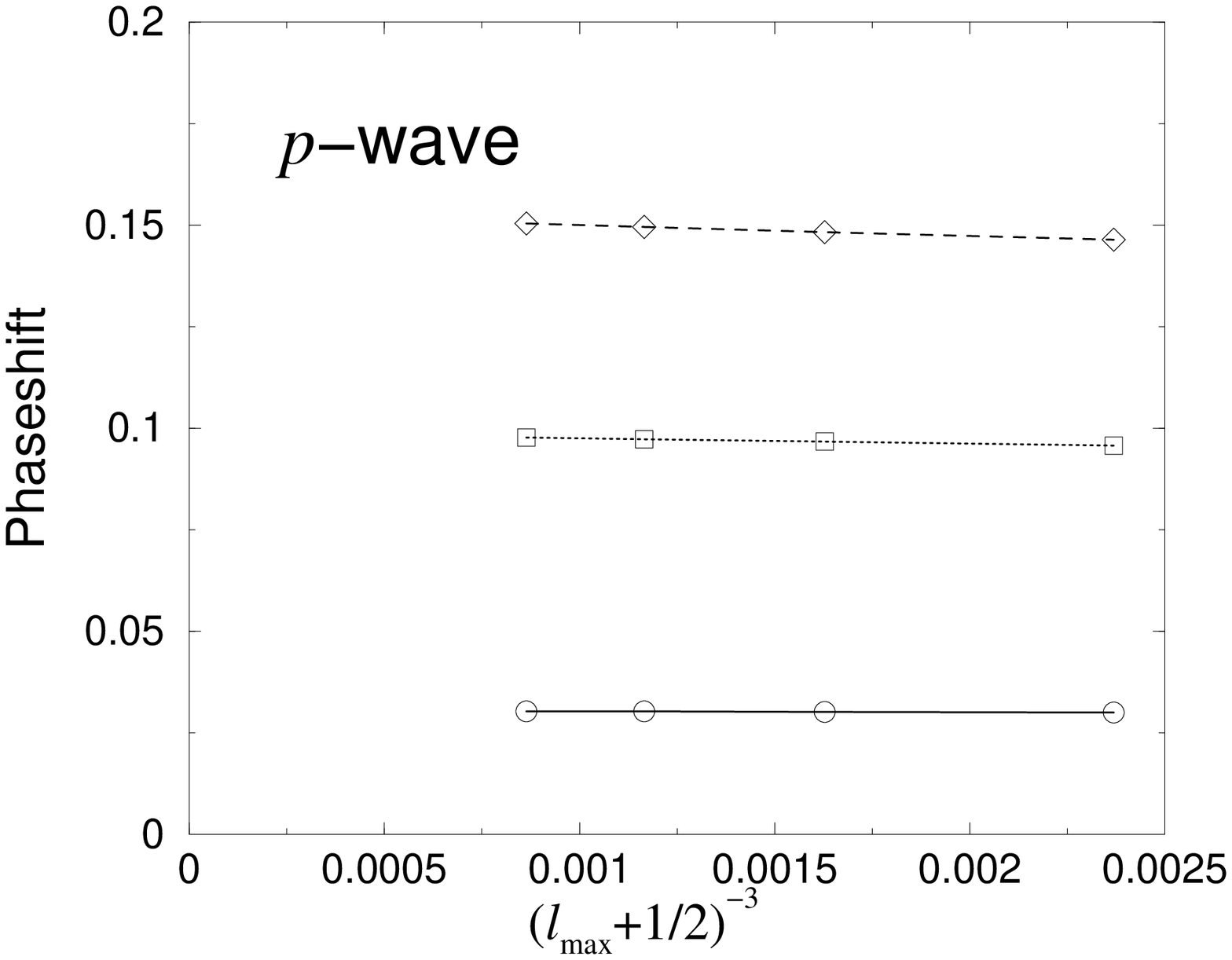}

\includegraphics[width=8cm]{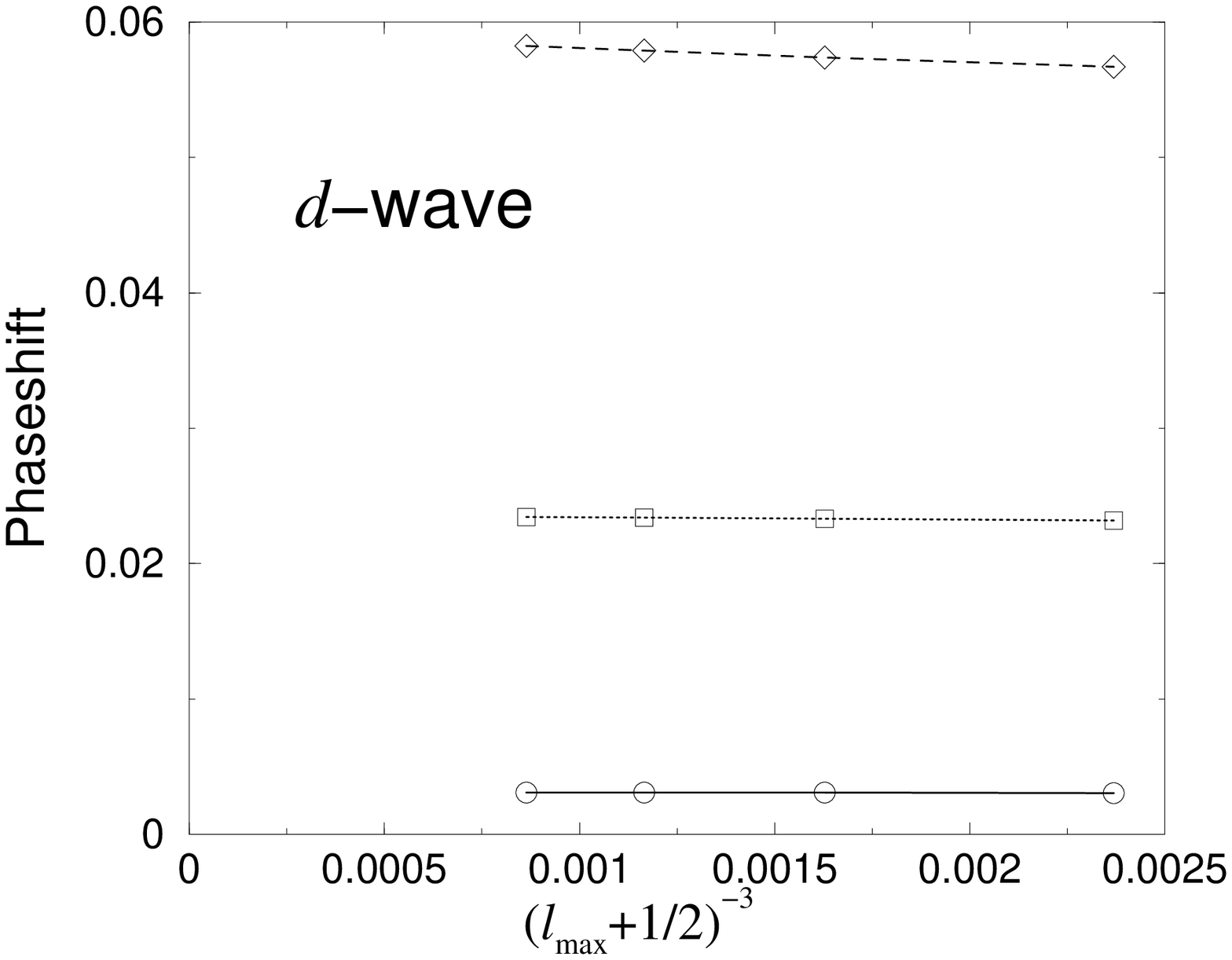}
\caption{Convergence of the $s$-, $p$- and $d$-wave positron scattering
phaseshifts on hydrogen with respect to the maximal orbital angular momentum
$l_{\rm max}$ for $R=15$~a.u. Circles, $k=0.2$~a.u.; squares,
$k=0.4$~a.u.; diamonds, $k=0.6$~a.u.}
\label{fig:ex_phase}
\end{figure}

\begin{figure}[ht]
\includegraphics[width=8cm]{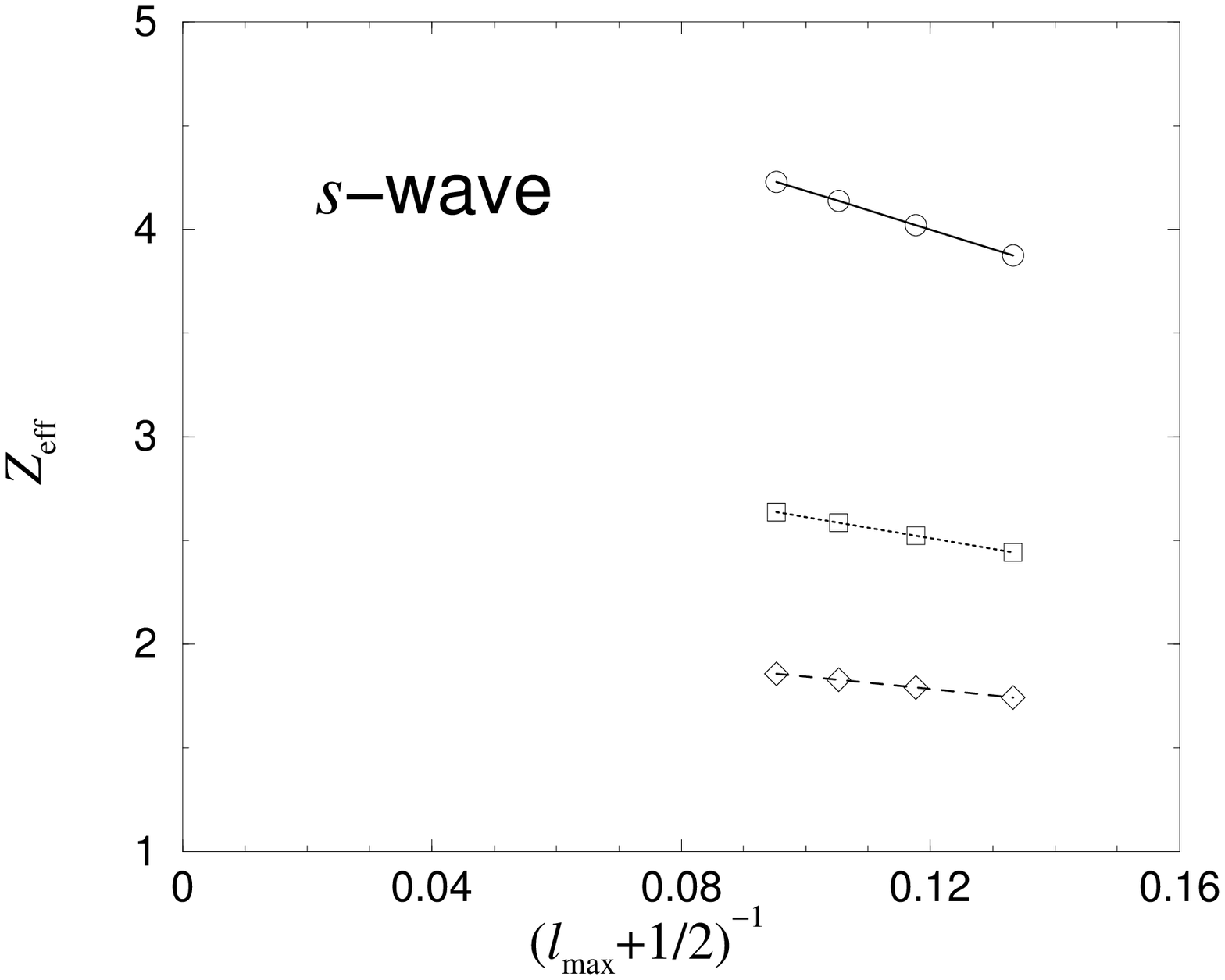}
\includegraphics[width=8cm]{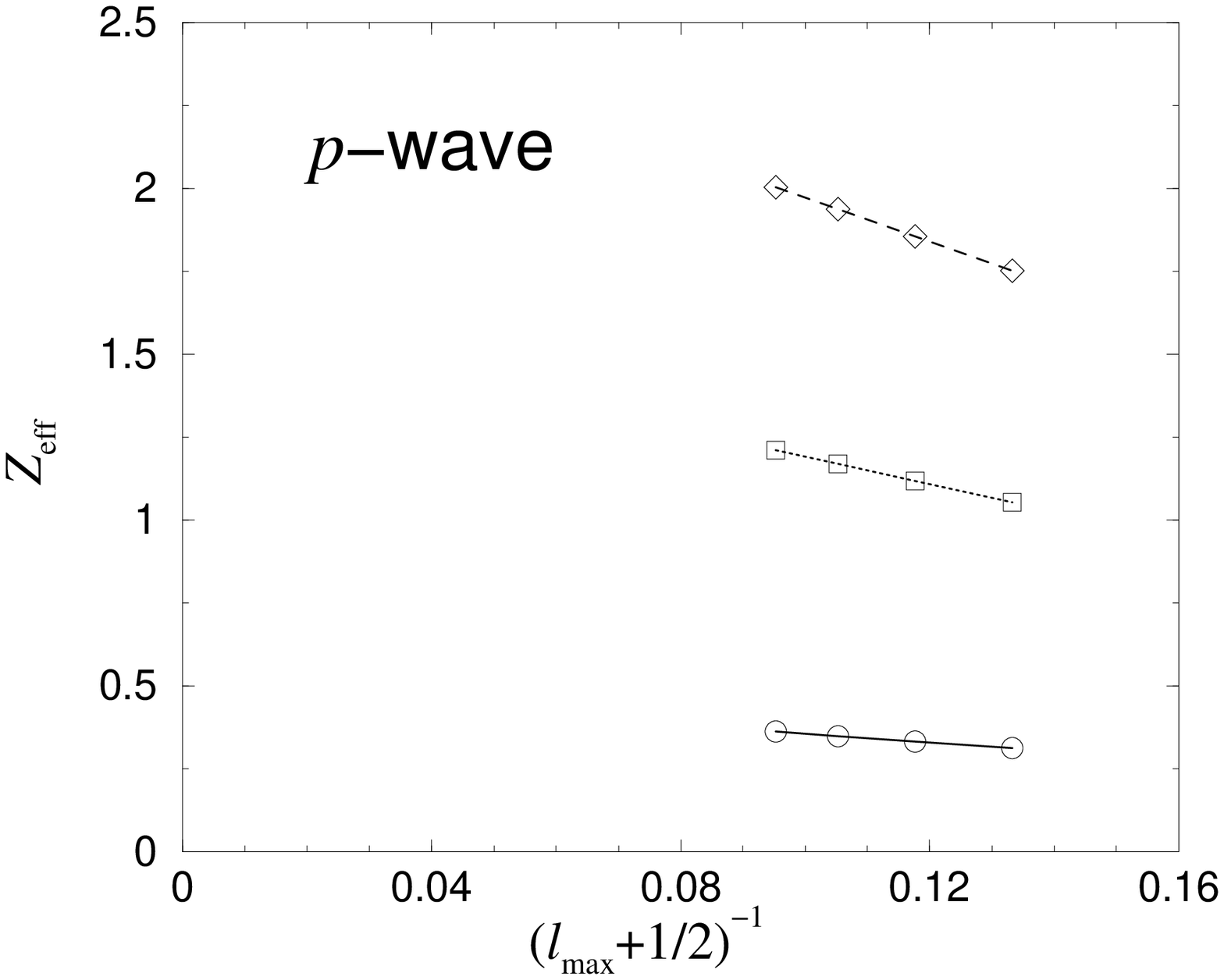}

\includegraphics[width=8cm]{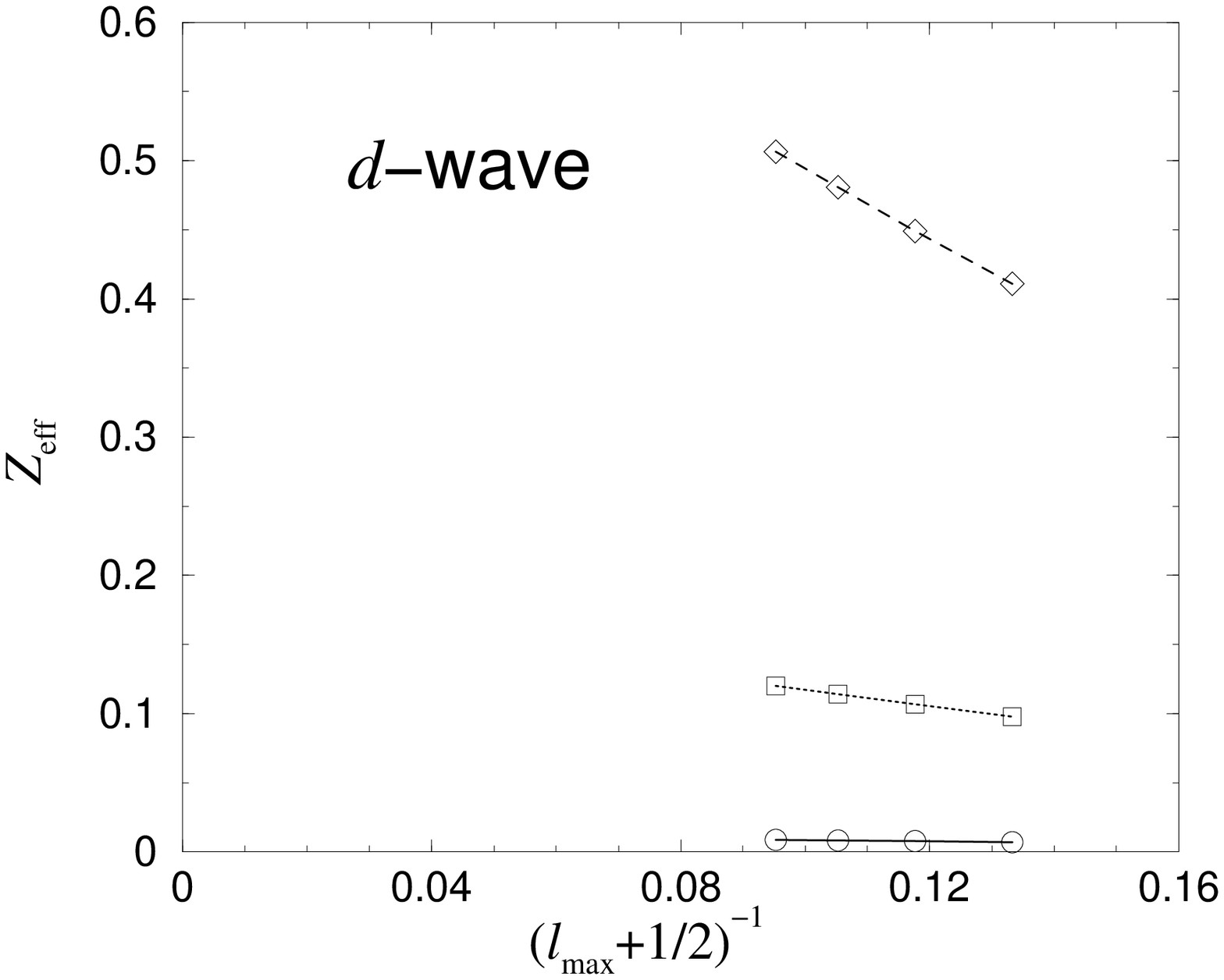}
\caption{Convergence of the $s$-, $p$- and $d$-wave contributions to
$Z_{\rm eff}$ for positron annihilation on hydrogen with respect to the
maximal orbital angular momentum $l_{\rm max}$ for $R=15$~a.u. Circles,
$k=0.2$~a.u.; squares, $k=0.4$~a.u.; diamonds, $k=0.6$~a.u.}
\label{fig:ex_Zeff}
\end{figure}

To extrapolate the scattering phaseshifts and $Z_{\rm eff}$
to $l_{\rm max}\rightarrow \infty $, as per Eqs. (\ref{eq:asym3})
and (\ref{eq:asym4}), the diagrams are evaluated for a range of maximal
orbital angular momenta, $l_{\rm max}=7$--10.
This procedure is illustrated by Fig. \ref{fig:ex_phase}
for the phaseshifts and Fig. \ref{fig:ex_Zeff} for $Z_{\rm eff}$, for the
$s$-, $p$- and $d$-wave incident positron with momenta $k=0.2$, $0.4$, and
$0.6$~a.u.
 
Figures \ref{fig:ex_phase} and \ref{fig:ex_Zeff} show that the calculations
have converged to the regime in which the asymptotic formulae (\ref{eq:asym3})
for $\delta _l$ and (\ref{eq:asym4}) for $Z_{\rm eff}$, may be applied.
The graphs also illustrate the point that the inclusion of high orbital
angular momenta and extrapolation to $l_{\rm max}\rightarrow \infty $ is much
more important in the calculations of annihilation, compared with scattering.
For the positron momenta and partial waves shown, between 15\% and 30\% of the
final value of $Z_{\rm eff}$ is due to such extrapolation. Quantitatively,
this contribution can be characterised by the ratio $B/Z_{\rm eff}$,
see Eq. (\ref{eq:asym4}), given in Table \ref{tab:BZeff}. Its increase
with the positron angular momentum may be related to the greater
role of the correlation corrections to the annihilation vertex in higher
positron partial partial waves (see below).

\begin{table}
\caption{Values of $B/Z_{\rm eff}$ which characterise the dependence
of calculated $Z_{\rm eff}$ on $l_{\rm max}$.\label{tab:BZeff}}
\begin{ruledtabular}
\begin{tabular}{cccc}
Momentum & \multicolumn{3}{c}{Partial wave}\\
 (a.u.) & $s$ & $p$ & $d$ \\
\hline
$0.2$ & 1.82 & 2.63 & 3.33 \\
$0.4$ & 1.62 & 2.58 & 3.32 \\
$0.6$ & 1.41 & 2.51 & 3.37 \\
\end{tabular}
\end{ruledtabular}
\end{table}

\begin{figure}[ht]
\includegraphics[width=8cm]{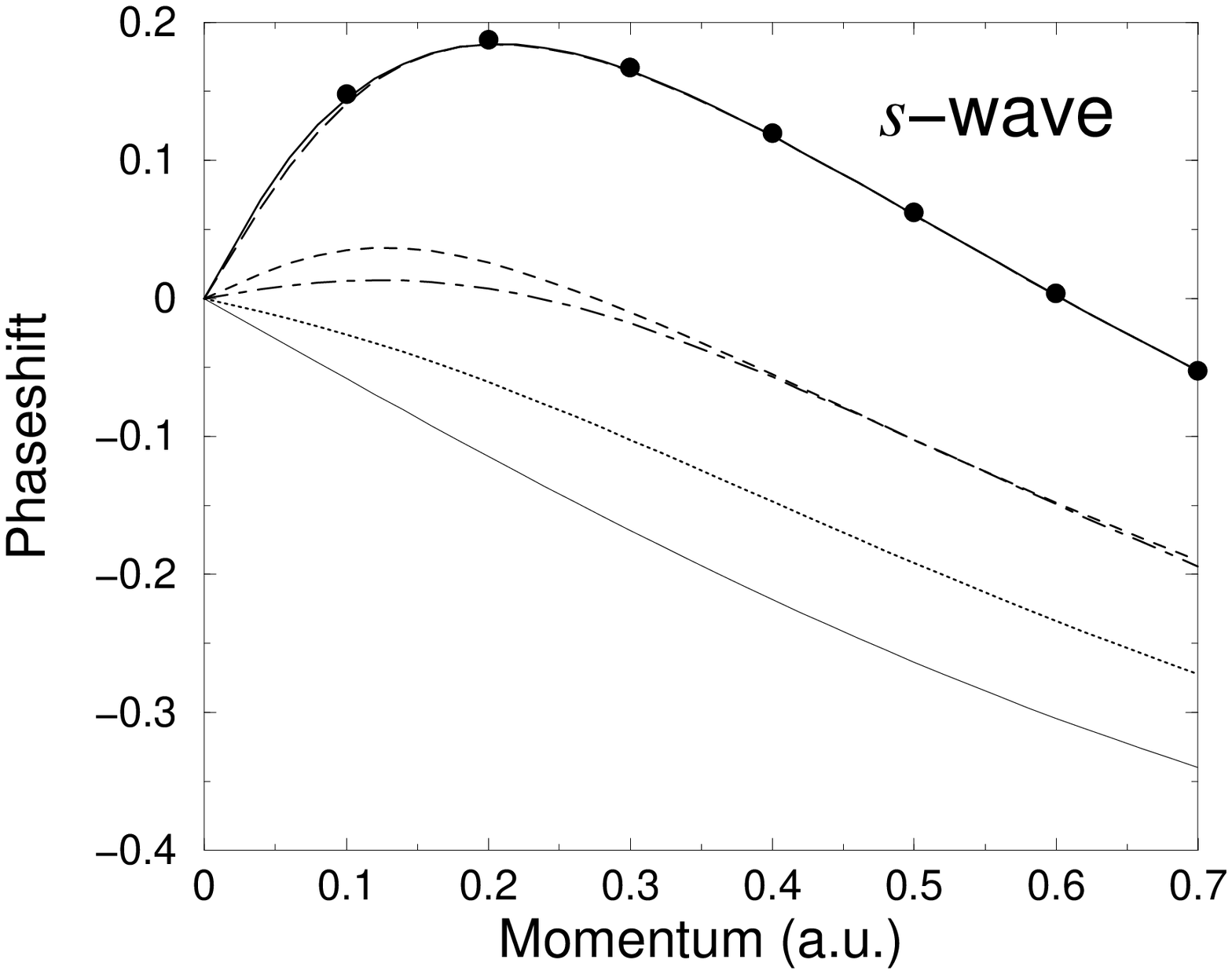}
\includegraphics[width=8cm]{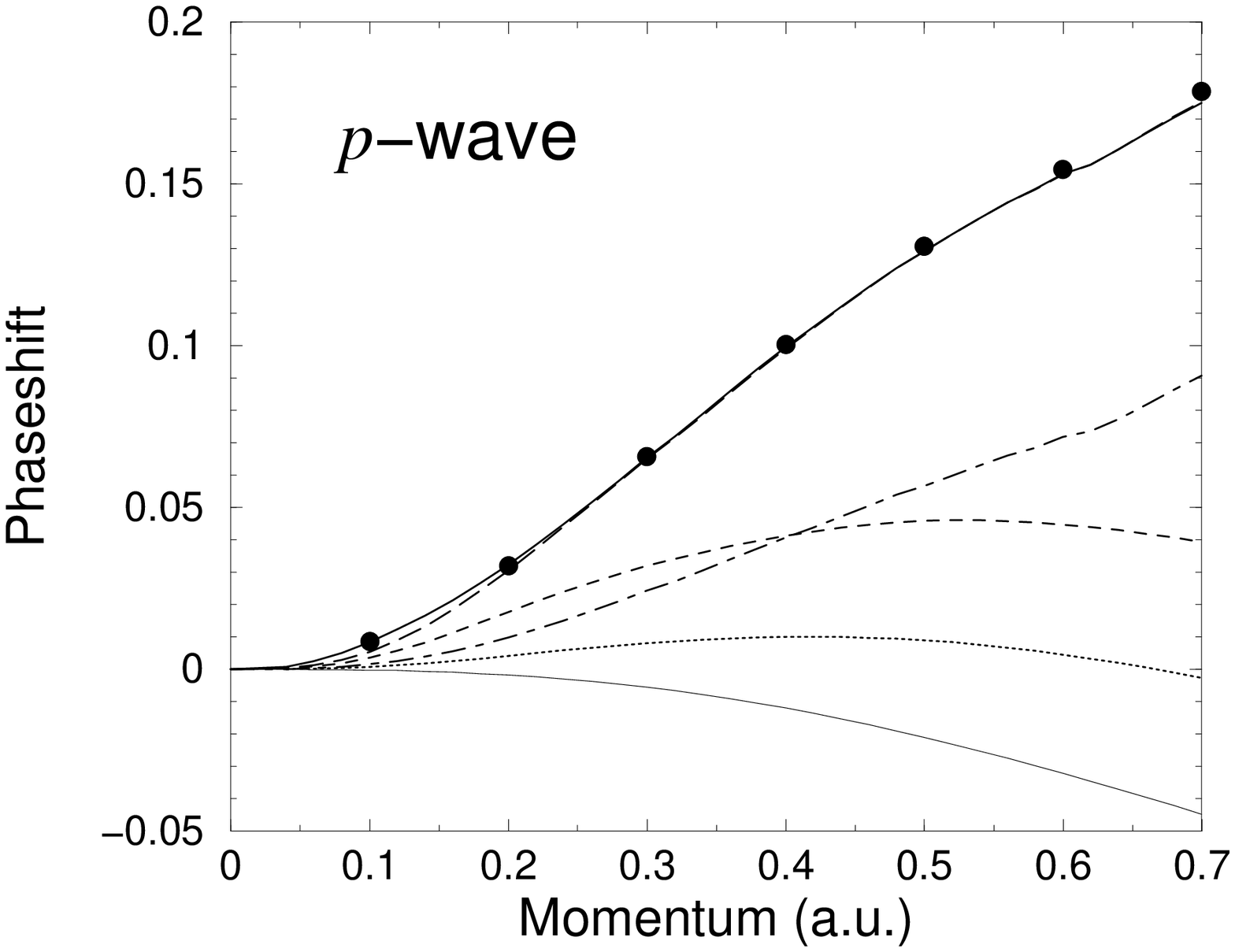}

\includegraphics[width=8cm]{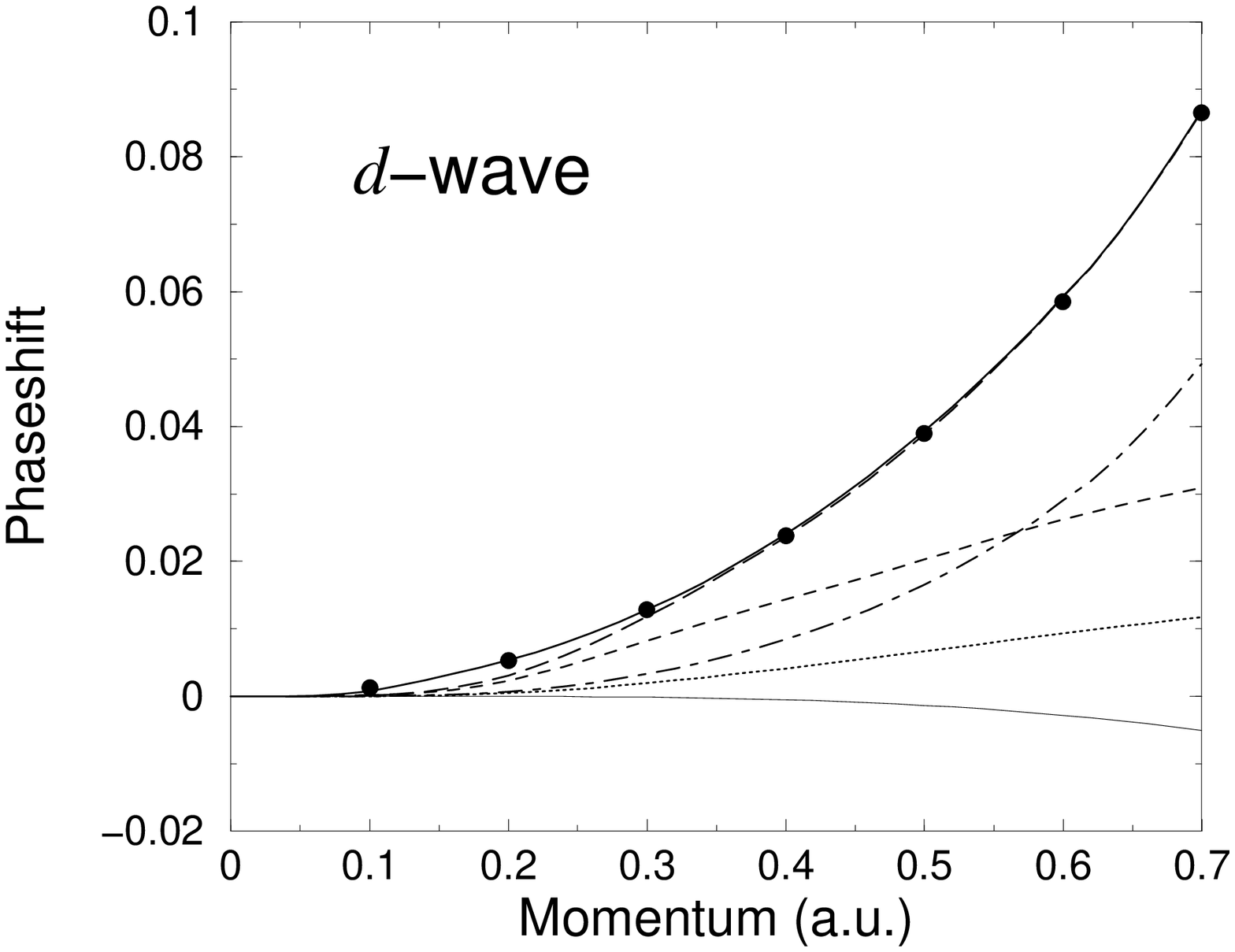}
\caption{Positron-hydrogen $s$, $p$ and $d$-wave scattering phaseshifts:
long-dashed curve, many-body theory ($R=15$~a.u.);
solid curve, many-body theory ($R=30$~a.u.); circles,
variational calculation \protect\cite{Humberston:97}. Thin solid curve,
static approximation; dashed curve, $\Sigma^{(2)}$; dot-dashed curve,
$\Sigma^{(\Gamma)}$; dotted curve, $\Sigma^{(\Gamma)}$ obtained with
$\Gamma=V$ (all $R=15$).}
\label{fig:phah}
\end{figure}

Figure \ref{fig:phah} shows the $s$-, $p$- and $d$-wave phaseshifts
for the total correlation potential (\ref{eq:sigtot}). They are in very good
agreement with those from an accurate variational calculation (see
Ref. \cite{Humberston:97}), the discrepancy being of order $10^{-3}$~rad.
The values obtained with $R=15$ and $R=30$~a.u. are almost
indistinguishable, except at low positron momenta. Here the results for
$R=30$ are superior to those for $R=15$. The larger cavity size allows
for a better account of the long range $-\alpha/2r^4$ tail in the
polarisation potential. 

Examining the phaseshifts allows us to compare the relative sizes
of the polarisation and Ps-formation contributions to the
correlation potential (\ref{eq:sigtot}).
The static positron-atom potential is repulsive, resulting in negative
values of the phase shifts (thin solid curves). The inclusion of
$\Sigma $, i.e., correlations, makes the low-energy phase shifts positive.
Dashed curves in Fig. \ref{fig:phah} show the phaseshifts obtained by
including only the 2nd-order diagram $\Sigma^{(2)}$ (polarisation), while
dot-dashed curves are those obtained with $\Sigma^{(\Gamma)}$ alone
(virtual Ps-formation). We see that none of these results is close to
the phaseshift obtained with the full $\Sigma $. This means that neither
contribution dominates the correlation potential, and the inclusion of both
polarisation and virtual Ps-formation effects is essential for solving the
positron-atom problem. Of course, any calculation which produces accurate
positron-hydrogen phaseshifts contains these contributions implicitly.
The advantage of the many-body theory approach is that one can separate
them, and get a better insight into the physics of the system.

To illustrate the non-perturbative nature of virtual Ps formation
we have also performed calculations that include the vertex function only to
1st order, $\Gamma _E=V$, in $\Sigma^{(\Gamma )}$ (dotted curves in
Fig. \ref{fig:phah}). This approximation accounts only for about 50\% of the
total vertex function contribution. Note that the higher-order terms in
$\Gamma $ become even more important close to the Ps formation threshold
($k\approx 0.7$) in $p$ and $d$ waves. This is related to the virtual Ps
becoming more ``real'' close to the threshold. 

\begin{figure}[ht]
\includegraphics[width=10cm]{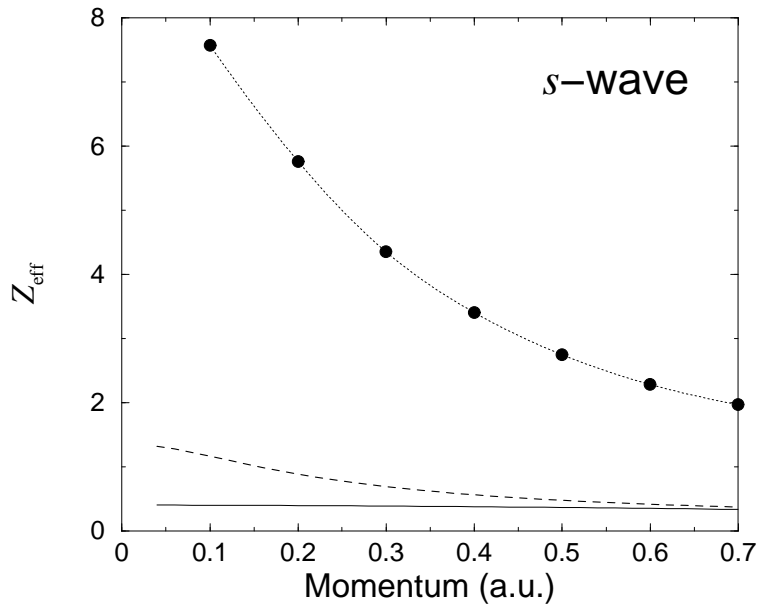}
\caption{Contribution of the 0th order diagram,
Fig. \protect{\ref{fig:Zeff1}}(a), to the positron-hydrogen $s$-wave
annihilation parameter $Z_{\rm eff}$, calculated with the
positron wavefunction in the static atomic potential (solid curve),
and with the Dyson orbital (dashed curve). Circles connected by dotted curve
are the accurate results of Refs. \protect\cite{VanReeth:98}.}
\label{fig:zz1}
\end{figure}

We now turn to positron annihilation. Having solved the scattering problem
accurately with the full $\Sigma $, we are now in possession of the best
(quasiparticle) positron wavefunction, the Dyson orbital.
Before using it in all annihilation diagrams, let us first
look at the effect of the Dyson orbital on the 0th-order diagram,
Fig. \ref{fig:Zeff1}(a), for the $s$-wave $Z_{\rm eff}$. Figure \ref{fig:zz1}
shows that the 0th-order contribution, Eq. (\ref{eq:Zeff0}), evaluated with
the positron wavefunction in the static atomic potential gives values
up to 20 times smaller than the accurate variational results
\cite{VanReeth:98}. This situation is similar to that in positron annihilation
on noble-gas atoms, where Eq. (\ref{eq:Zeff0}) evaluated with the static
(HF) positron wavefunction underestimates experimental
$Z_{\rm eff}$ by a factor of $10^1$--$10^3$ at low positron energies
\cite{Dzuba:93,Dzuba:96}. It is natural that the use of the Dyson orbital,
which is ``aware'' of the positron-atom attraction, in Eq. (\ref{eq:Zeff0}),
leads to increased $Z_{\rm eff}$, and introduces a correct momentum dependence
(dashed curve in Fig. \ref{fig:zz1}). This latter fact is in agreement with
the general understanding of the origins of the energy dependence 
and enhancement of $Z_{\rm eff}$ at low energies, and their relation to
positron-atom virtual states \cite{Dzuba:93,Dzuba:96,Gribakin:00}.
However, the absolute values obtained are still about 5 times lower
than the benchmark.

\begin{figure}[pht]
\includegraphics[width=8cm]{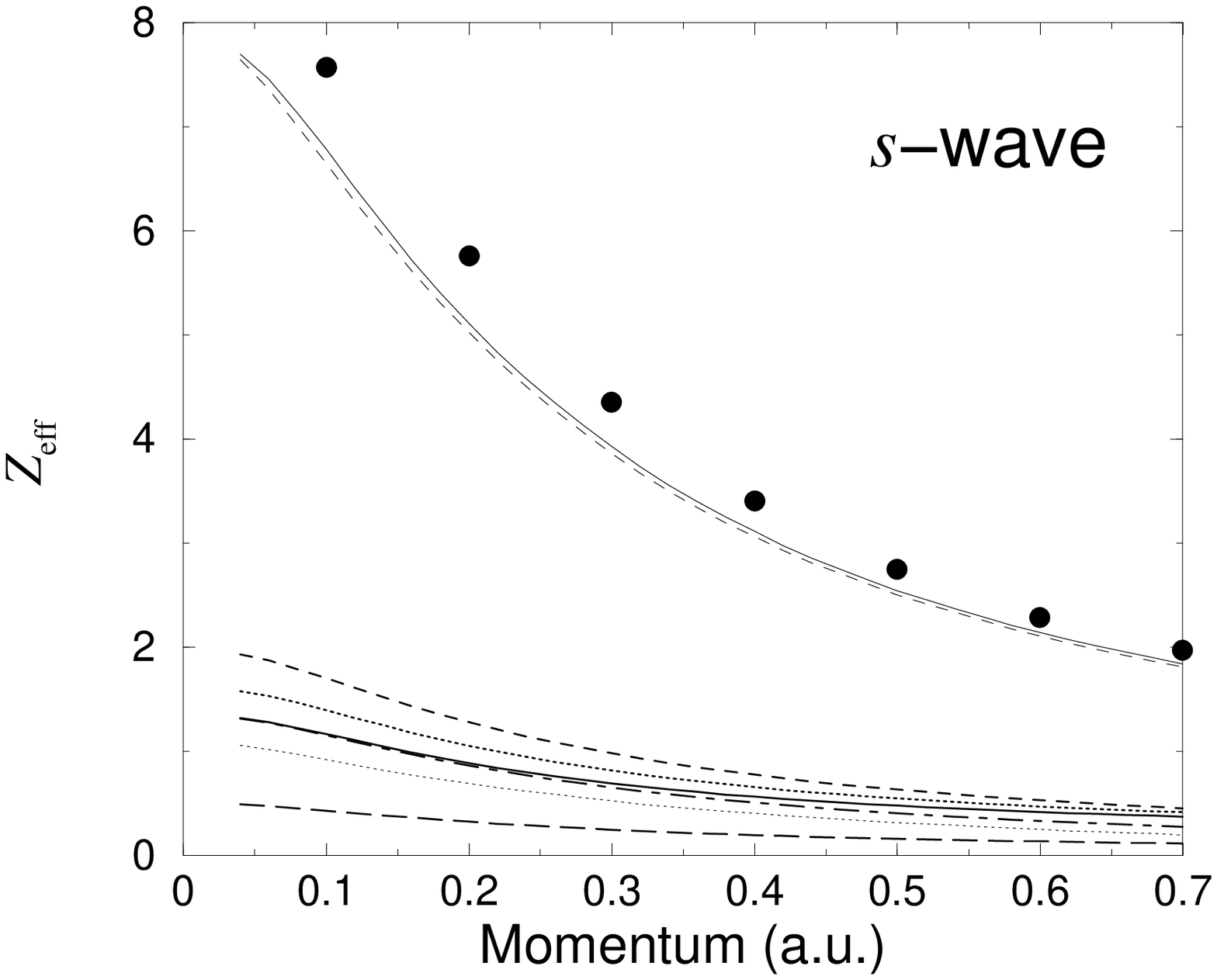}
\includegraphics[width=8cm]{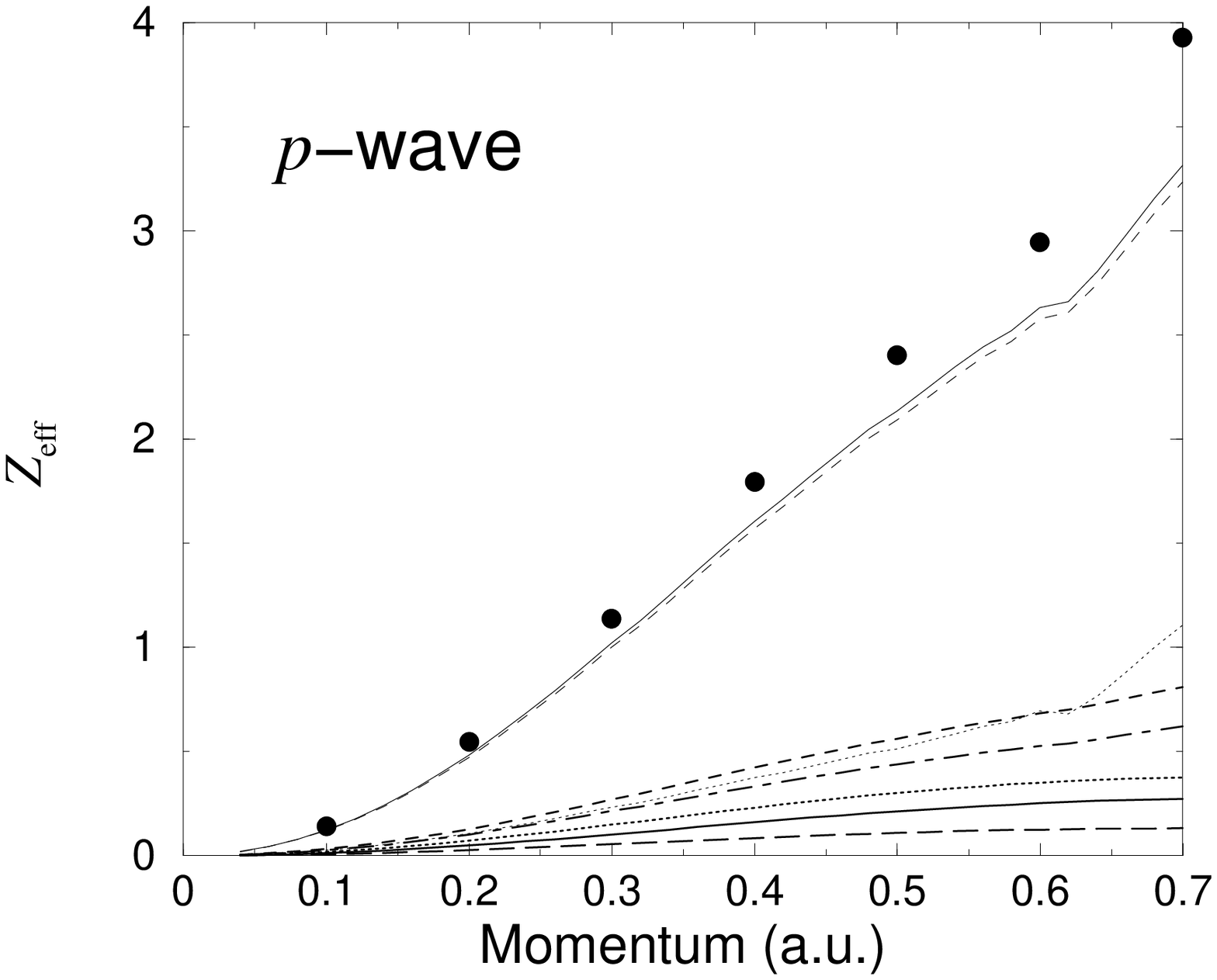}

\includegraphics[width=8cm]{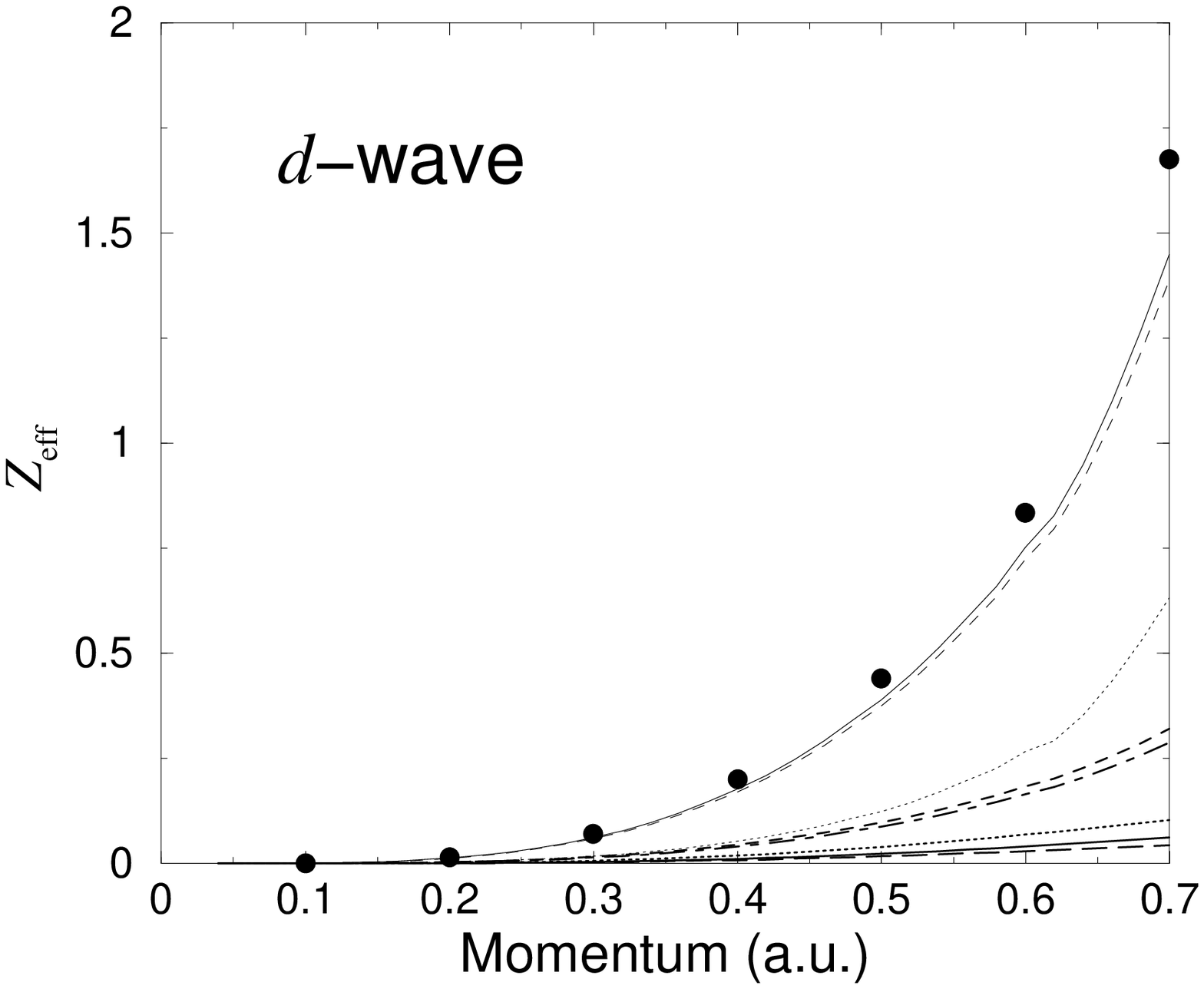}
\caption{Annihilation parameter $Z_{\rm eff}$ for the $s$-, $p$-, and $d$-wave
positron on hydrogen. Contributions of individual diagrams from
Fig. \protect{\ref{fig:Zeff1}} are: thick solid line, diagram (a);
thick dotted line, (b); long dashed line, (c);
thick dashed line, (d); dot-dashed line, (e);
thin dotted line, (f) (all for $R=15$~a.u.).  Thin solid line is the total
for $R=15$~a.u., thin dashed line, total for $R=30$~a.u.
Circles are the results of Ref. \protect\cite{VanReeth:98}.}
\label{fig:zmany}
\end{figure}

The remaining 80\% come from the nonlocal corrections to the annihilation
vertex, diagrams (b)--(f) in Fig. \ref{fig:Zeff1}. The contributions of
all the diagrams evaluated using the positron Dyson orbitals, and the total
$Z_{\rm eff}$ are shown in Fig. \ref{fig:zmany} for the positron $s$, $p$ and
$d$ partial waves.

The difficulty of calculating the vertex corrections to $Z_{\rm eff}$
accurately is evident from these graphs, as all the diagrams in
Fig. \ref{fig:Zeff1} contribute significantly. The higher-order diagrams
containing the vertex function, are close to or greater than
the lower-order diagrams. Note that all contributions have a similar
dependence on the positron momentum. It is driven by the momentum
dependence of the positron Dyson orbitals (external lines
in the diagrams), as discussed in Sec. \ref{subsec:ann}.

Figure \ref{fig:zmany} shows that for $p$ and $d$ waves, the contribution
of the diagram Fig. \ref{fig:Zeff1}(f) grows rapidly and becomes
largest towards the Ps formation threshold. This diagram describes
annihilation inside the virtual Ps formation, which has a vigorous energy
dependence close to threshold \cite{threshold}. This may have a bearing on
the kinks in $Z_{\rm eff}$ visible at $k\approx 0.6$~a.u. for the $p$ and $d$
waves. Although they may be a numerical artifact, an indication of an
inflection point is also present in the accurate $p$-wave results of
Ref. \cite{VanReeth:98}.

\begin{figure}[ht]
\includegraphics[width=10.0cm]{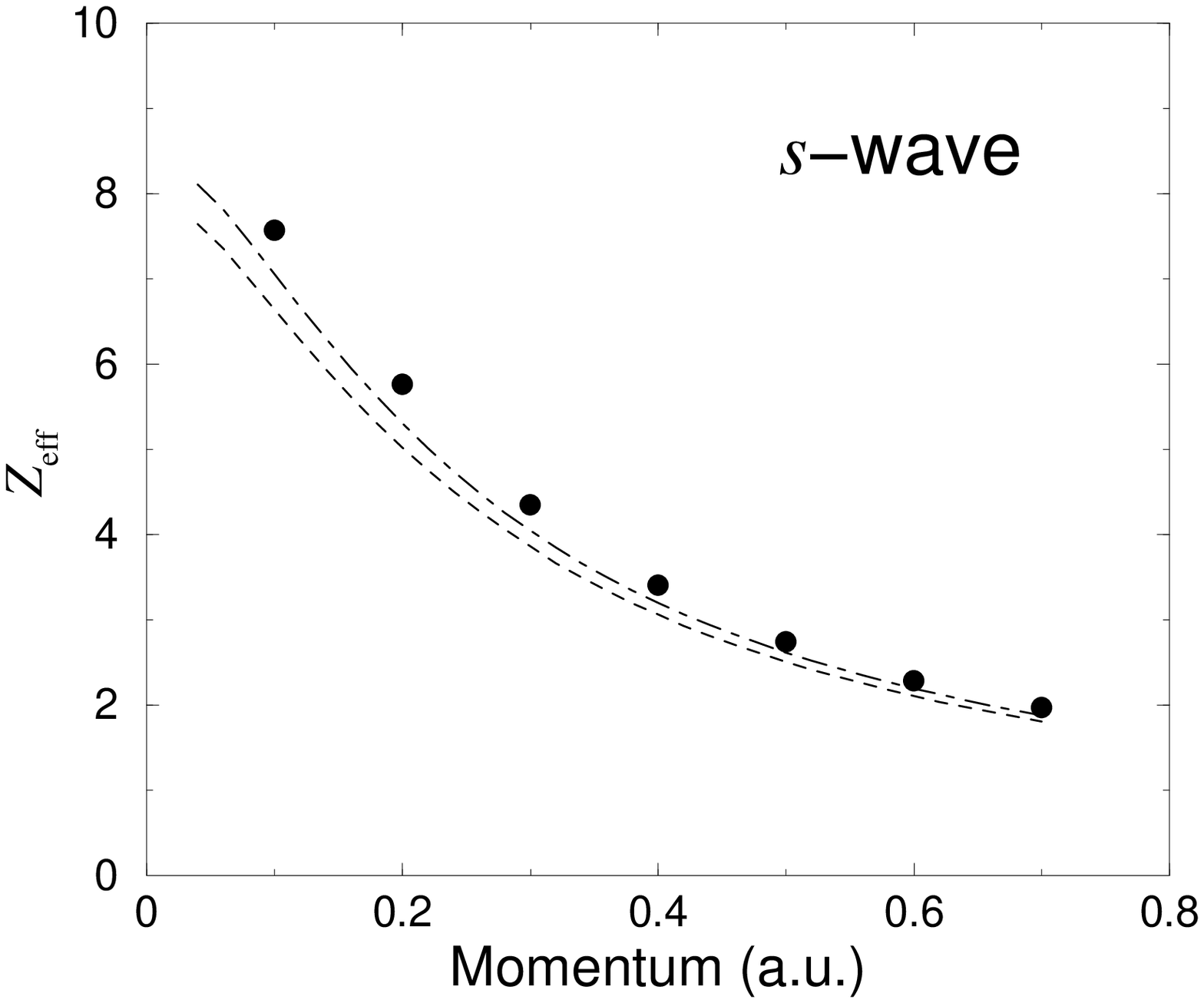}
\caption{Positron annihilation on hydrogen, $s$ wave.
Many-body theory results: dashed line, $R=30$~a.u., $n=40$, $k=6$, first 15
states used; dot-dashed line, $R=30$~a.u., $n=60$, $k=9$, first 23 states
used; circles, Ref. \protect\cite{VanReeth:98}.}
\label{fig:zmany2}
\end{figure}

The final results for $R=15$~a.u. are slightly higher than for $R=30$~a.u.
Indeed, annihilation takes place near the atom, and a denser knot
sequence for $R=15$ may provide a better description of small electron-positron
separations. Our final values compare well with the accurate positron-hydrogen
results obtained using a variational approach \cite{VanReeth:98}.
In that work the electron-positron distance was represented
explicitly in the calculation, while we use a single-centre expansion.
We believe that the remaining small discrepancy could be eliminated
by `pushing harder' the numerics in our approach.
Thus, Fig. \ref{fig:zmany2} shows that if we use $n=60$ B-splines of order
$k=9$ and include the first 23 basis states, the difference between our
$Z_{\rm eff}$ and the benchmark values is halved.

The main conclusion of this section is that the numerical implementation of the
many-body theory approach proposed in this paper works. For positron
collisions with hydrogen, where this approach is exact, the calculations
reproduce the best scattering phaseshifts, and yield good results
in the more difficult annihilation problem.


\section{Correlation contribution to the annihilation vertex.}

The ability of a many-body theory to describe correlation corrections to
the annihilation vertex may give some insight into the role such
corrections play in positron annihilation with matter.
Thus, in theoretical studies of positron annihilation in condensed matter,
one usually starts by performing a calculation for the ground state electron
and positron densities. The annihilation rate is then found via an equation
of the form (see, e.g., \cite{Puska}),
\begin{equation}\label{eq:cond}
\lambda = \pi r_0^2c\int \rho_e({\bf r})\rho_p({\bf r})
\gamma(\rho_e,\rho_p){\rm d}{\bf r},
\end{equation}
where $\rho_p({\bf r})$ and $\rho_e({\bf r})$ are the positron and electron
densities and $\gamma(\rho_e,\rho_p)$ is the so-called enhancement
factor. It is introduced to account for the Coulomb attraction in the
annihilating electron-positron pair. It has long been known that the
independent-particle approximation ($\gamma =1$) underestimates the
annihilation rates by several times \cite{Ferrell:56}, and a number of
semi-empirical and interpolation forms of $\gamma(\rho_e,\rho_p)$ have been
suggested (see, e.g., Ref. \cite{Barb} and references therein).

A comparison between Eq. (\ref{eq:cond}) with $\gamma =1$ and the many-body
theory expression (\ref{eq:nonloc}) for the annihilation rate, shows that
the former corresponds to the 0th-order term in $Z_{\rm eff}$,
Eq. (\ref{eq:Zeff0})
\cite{correction}. As we have seen in Sec. \ref{sec:res} (Fig. \ref{fig:zz1}),
$Z_{\rm eff}^{(0)}$ does underestimate the annihilation rate in hydrogen
by a factor of 5, even when the best positron wavefunction is used. On the
other hand, the correlation correction to the annihilation rate [2nd term
in Eq. (\ref{eq:nonloc})] does not have the form of Eq. (\ref{eq:cond}).
This correction depends on the positron wavefunction at two different points,
${\bf r}$ and ${\bf r}'$. Therefore, any local expression like
Eq. (\ref{eq:cond}) is necessarily an approximation.

\begin{figure}[ht]
\includegraphics[width=10.0cm]{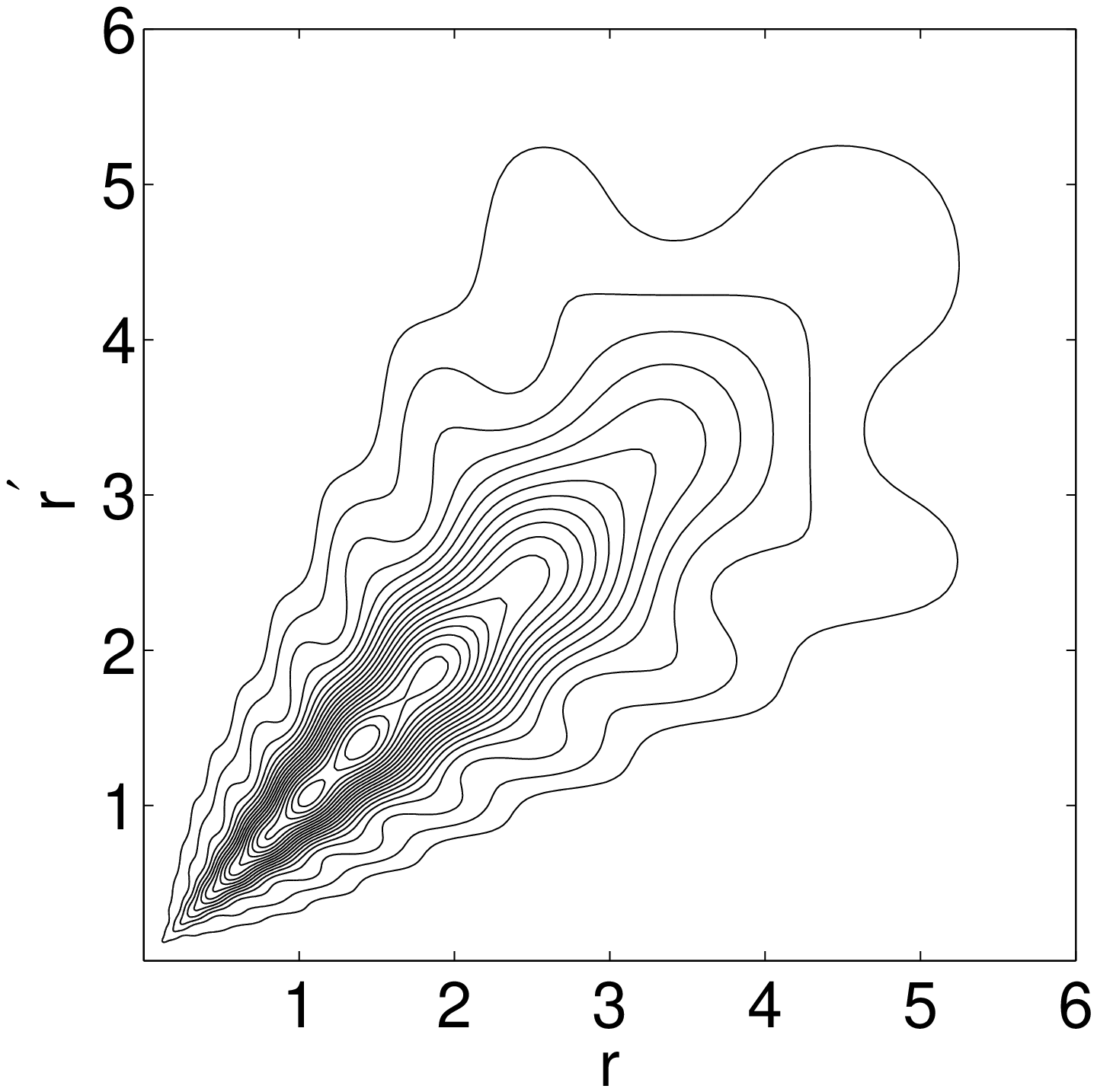}
\caption{Radial dependence of the integrand (\protect\ref{eq:integ}) of the
correlation correction for the annihilation of the $s$-wave positron with
momentum $k=0.06$~a.u. on hydrogen. The ``ripples'' is an
artifact of the reconstruction of $\Delta _\eps ^{(l)}(r,r')$
from the matrix element $\langle i|\Delta _\eps |j\rangle $.} 
\label{fig:contour}
\end{figure}

To illustrate this point, a contour plot in Fig. \ref{fig:contour}
shows the radial part of the integrand of the nonlocal term,
\begin{equation}\label{eq:integ}
P_{\eps l}(r)\Delta _{\eps}^{(l)}(r, r^{\prime})P_{\eps l}(r^{\prime}) ,
\end{equation}
for the $s$-wave ($l=0$) positron annihilation on hydrogen at $k=0.06$~a.u. 
The plot confirms that the correlation contribution to the annihilation
vertex is localised near the atom. Its maximum at
$r=r^{\prime}\approx 1.5$~a.u. compares well with the radius of the
hydrogen atom, and the ridge-like structure indicates that the
``nonlocality'' is limited to $|r-r^{\prime}|\sim 1$~a.u.

The overall size of the (nonlocal) correction to the annihilation
vertex can be characterised by the average enhancement factor
$\bar{\gamma}$,
\begin{equation}\label{eq:bar_gam}
\bar{\gamma}=1+\frac{\int \psi_ \eps ^*({\bf r})
\Delta _\eps ({\bf r},{\bf r}')\psi _\eps ({\bf r}')
d{\bf r}d{\bf r}'}
{\int \sum_n |\varphi _n({\bf r})|^2|\psi_{\eps}({\bf r})|^2d{\bf r}}.
\end{equation}
We defined this factor in such a way that when used in place
of $\gamma $ in Eq. (\ref{eq:cond}),
together with $\rho _e({\bf r})=\sum_n |\varphi _n({\bf r})|^2$ and
$\rho _p({\bf r})=|\psi_{\eps}({\bf r})|^2$, it would reproduce correct values
of the positron-atom annihilation rate.

The quantity $\bar{\gamma}$ can also be defined as the ratio of the total
$Z_{\rm eff}$ to the value obtained from the 0th-order diagram,
$Z_{\rm eff}^{(0)}$, Fig. \ref{fig:Zeff1}(a). Because of the weak energy
dependence of $\Delta _\eps $ (see Sec. \ref{subsec:ann}), this ratio should
not depend strongly on the energy of the incident positron. This also implies
that $\bar{\gamma}$ will be relatively insensitive to the wavefunction used
to describe the incident positron. In particular, the use of either HF
or Dyson wavefunctions $\psi _\eps $ for the incident positron should yield
close values of $\bar \gamma $.

\begin{figure}[ht]
\includegraphics[width=10cm]{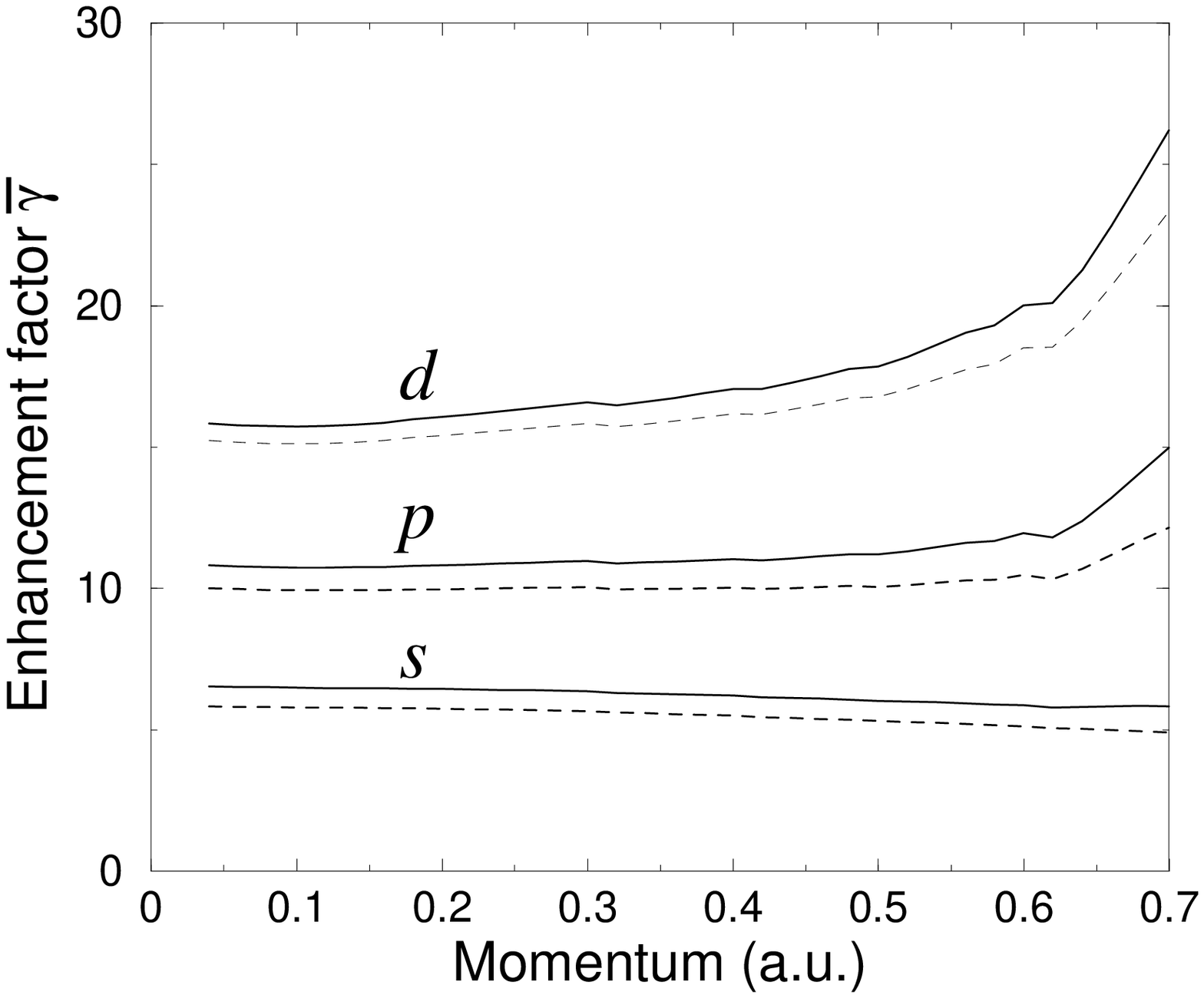}
\caption{Ratio $\bar{\gamma}$ of the total $Z_{\rm eff}$ to
$Z_{\rm eff}^{(0)}$ for positron annihilation on hydrogen.
Values obtained with positron wavefunction in the
static approximation are shown by solid curves, and those with the
Dyson orbitals, by dashed curves.}
\label{fig:ratioz}
\end{figure}

As a test, this ratio was evaluated for the annihilation of the $s$, $p$ and
$d$-wave positrons on hydrogen, see Fig. \ref{fig:ratioz}. 
As expected, the values of $\bar{\gamma}$ depend weakly on the incident
positron energy, except when the Ps formation threshold is approached.
This confirms the earlier observation that the momentum dependence
of various contributions to $Z_{\rm eff}$ in Fig. \ref{fig:zmany}
is approximately the same. The
results obtained within the static approximation and with the Dyson orbitals
are also very close to each other, even though the absolute values of
$Z_{\rm eff}$ obtained in the two approximations are very different
(see, e.g., Fig. \ref{fig:zz1}). One cannot help noticing that the enhancement
due to annihilation vertex
corrections increases with the angular momentum of the positron. For larger
$l$, the penetration of the positron into the electron-rich regions of the
atom is suppressed by the centrifugal barrier, and the effect of
``pulling the electron out'' (i.e., virtual Ps formation), described by the
correlation corrections, becomes more important.

Equation (\ref{eq:cond}) is local, whereas equation (\ref{eq:nonloc}) contains
a non-local term. In order to compare the enhancement of $Z_{\rm eff}$
due to this term with the enhancement factors used in condensed matter
calculations, we need to ``localise'' the contribution of
$\Delta _\eps ({\bf r},{\bf r}')$. Let us do this by introducing an
effective `electron correlation density' $\tilde{\rho}_e(r)$ via the relation
\begin{equation}\label{eq:rhoe}
\int \!\!\!\int P_{\eps l}(r)\Delta _\eps ^{(l)}(r,r')P_{\eps l}(r')drdr'\equiv
\int \tilde{\rho}_e(r)P_{\eps l}^2(r)dr,
\end{equation}
for positron annihilation in the $l$th partial wave. This allows us to define
an effective enhancement factor $\gamma _e(r)$ through
\begin{equation}\label{eq:gameff}
\rho _e(r)+\tilde \rho _e(r)\equiv \gamma _e(r)\rho _e(r) ,
\end{equation}
which is equivalent to
\begin{equation}\label{eq:gameff1}
\gamma _e(r)=1+\tilde \rho _e(r)/\rho _e(r) ,
\end{equation}
where $\rho_e(r)=P_{1s}^2(r)/4\pi r^2$, and $P_{1s}=2re^{-r}$ for
hydrogen. Equations (\ref{eq:rhoe}) and (\ref{eq:gameff1}) guarantee
that if we use $\gamma _e (r)$ in Eq. (\ref{eq:cond}), correct annihilation
rates will be recovered.

Equation (\ref{eq:rhoe}) does not define $\tilde \rho _e(r)$ uniquely. We
use two different methods to obtain it numerically.
The first one states,
\begin{equation}\label{eq:rhoe1}
\tilde{\rho}_e^{[1]}(r) = \frac{\int \Delta_{\eps}^{(l)}(r,r^{\prime})
P_{\eps l}(r^{\prime})dr^{\prime}}{P_{\eps l}(r)}
\end{equation}
which ensures that $\tilde{\rho}_e^{[1]}(r)$ satisfies
Eq. (\ref{eq:rhoe}) exactly. However, it has a disadvantage in that it may
have unphysical poles at the zeroes of the positron wavefunction. A second
method involves calculating
\begin{equation}\label{eq:rhoe2}
\tilde{\rho}_e^{[2]}(r) = \int_{-2r}^{2r}
\Delta_{\eps}^{(l)}(r+\sigma/2,r-\sigma/2) d \sigma ,
\end{equation}
which follows from Eq. (\ref{eq:rhoe}) if we change variables 
$r,\,r'$ to $r\pm \sigma /2$, and keep the lowest order
term in the Taylor expansion of $P_{\eps l}(r+\sigma/2)P_{\eps l}(r-\sigma/2)$.
This approximation may not be accurate for small $r$,
where the positron wavefunction varies rapidly, but should be correct for
larger $r$ values, where the peaking
of $\Delta _\eps ^{(l)}(r,r')$ at $r=r'$ (Fig. \ref{fig:contour}) means
that $P_{\eps l}(r)$ varies slowly on the scale of typical $\sigma $.

\begin{figure}[ht]
\includegraphics[width=10cm]{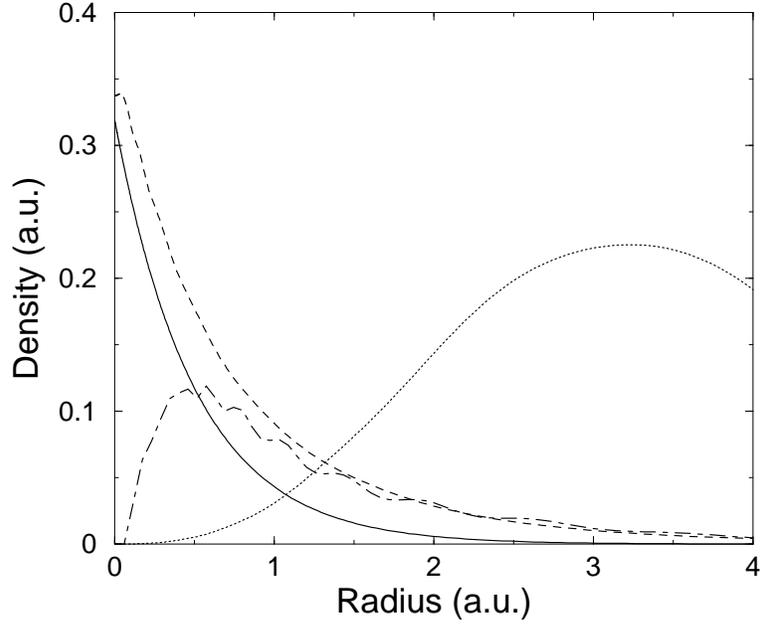}
\caption{Effective electron correlation densities $\tilde{\rho}_e^{[1]}(r)$
(dashed curve) and $\tilde{\rho}_e^{[2]}(r)$ (dot-dash curve) for the
annihilation of the $s$-wave positron with $k=0.5$ on hydrogen.
Shown for comparison are the ground-state electron density in hydrogen
$\rho_e(r)$ (solid curve), and the positron density $P_{\eps l}^2(r)$
(dotted curve, arbitrarily scaled).}
\label{fig:rhoe}
\end{figure}

Figure \ref{fig:rhoe} shows both electron correlation densities calculated
for the $s$-wave positron with momentum $k=0.5$ a.u. Apart from the 
small range of distances near the origin, the values of $\tilde \rho _e$
from the two methods are close, although $\tilde{\rho}_e^{[2]}$ shows
some numerical ``noise'' related to the reconstruction of
$\Delta _\eps ^{(l)}(r,r')$ from its matrix elements. A comparison with the
hydrogen ground-state electron density shows that the latter drops much faster
with the distance from the nucleus. In fact, $\tilde \rho _e$ is much greater
than $\rho _e$ at those $r$ where the positron density is large, in agreement
with the correlation contribution to $Z_{\rm eff}$ being five times
$Z_{\rm eff}^{(0)}$.

\begin{figure}[ht]
\includegraphics[width=10cm]{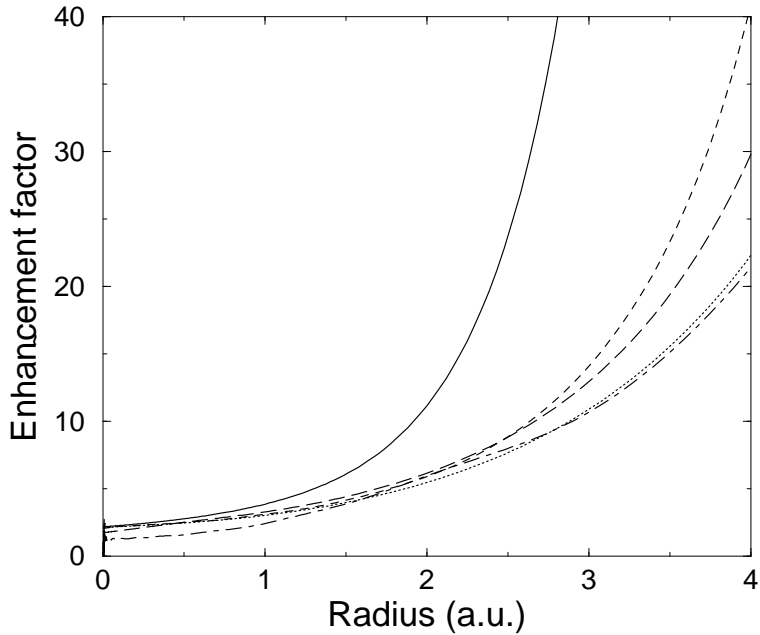}
\caption{Enhancement factors for positron annihilation on hydrogen.
Solid curve, $\gamma_{AP}$, Eq. (\protect\ref{eq:AP});
dotted line $\gamma _e$ for $s$-wave positrons at $k=0.06$; 
dashed line,$\gamma _e$ for $s$-wave positrons at $k=0.5$;
long-dashed line, $\gamma _e$ for $p$-wave positrons at $k=0.5$;
dot-dashed line, $\gamma _e$ for $d$-wave positrons at $k=0.5$.
}
\label{fig:gamma}
\end{figure}

The electron correlation densities obtained above allow us to calculate the
corresponding enhancement factors and compare them with a
parametrisation of $\gamma (\rho_e ,\rho_p)$ derived by Arponen and Pajanne
\cite{AP} for a positron in a homogeneous electron gas,
\begin{equation}\label{eq:AP}
\gamma_{\rm AP} = 1+1.23r_s-0.0742r_s^2+\frac{1}{6}r_s^3,
\end{equation}
where $r_s$ is a measure of the average distance between the electrons,
\begin{equation}\label{eq:rs}
r_s = \left(\frac{3}{4\pi\rho_e}\right)^{1/3} .
\end{equation}
This enhancement factor is plotted in Fig. \ref{fig:gamma} together
with the effective enhancement factors $\gamma _e(r)$ derived from
Eq. (\ref{eq:gameff1}) using $\tilde \rho _e ^{[1]}$, which is more
stable numerically than $\tilde \rho _e ^{[2]}$. Note that
unlike $\gamma _{\rm AP}$, the factor $\gamma _e$
obtained from our many-body theory approach, depends on the energy
and orbital angular momentum of the positron. The values shown in
Fig. \ref{fig:gamma} correspond to the momenta, $k=0.06$ ($s$ wave)
and 0.5 a.u. ($s$, $p$ and $d$ waves).

A feature common to all enhancement factors in
Fig. \ref{fig:gamma} is their rapid rise with the distance from the
nucleus. This increase is related to the drop of the electron
density, a relation which is explicit in Eq. (\ref{eq:AP})
for $\gamma_{\rm AP}$. At small distances, where the electron density is large,
$\gamma _{\rm AP}$ compares well with $\gamma _e$. However, at
larger distances, where the electron density is low, $\gamma _{\rm AP}$
is much greater than all of the $\gamma _e$. Such a discrepancy could
be expected, given that a homogeneous electron gas theory used to derive
$\gamma _{\rm AP}$ is more reliable in the high-density limit.
The exaggeration of the enhancement by $\gamma _{\rm AP}$ was reported
in Ref. \cite{Barb} where various forms of the enhancement factor were
tested by comparison with accurate annihilation rates for a number of
positron-atom bound systems. For Be and Mg (which have $I>6.8$ eV), the
values obtained using $\gamma _{\rm AP}$ overestimated the accurate
annihilation rates by factors of 5 and 2, respectively. 
It must be noted that the product $\rho _e\gamma _{\rm AP}$ remains
finite as $\rho _e\rightarrow 0$. This means that one cannot in principle
use it in Eq. (\ref{eq:cond}) for a continuous spectrum positron, since it
would yield infinite values of the annihilation rate  and $Z_{\rm eff}$.
For the same reason a much stronger overestimate observed in
Ref. \cite{Barb} with $\gamma _{\rm AP}$ for Be is a direct consequence
of the positron binding energy for Be being much smaller than for Mg.

Figure \ref{fig:gamma} confirms that $\gamma _e$ is energy-dependent.
It is greater for $k=0.5$~a.u. compared with $k=0.06$ ($s$ wave),
the higher momentum being closer to the Ps formation threshold.
Values of $\gamma _e$ derived for the $p$ and $d$ waves are similar to those
from the $s$ wave. Hence, the large differences between the
average enhancement factors in Fig. \ref{fig:ratioz} are due to
the effect of the centrifugal barrier on the positron wavefunction.
The distances which effectively contribute to $Z_{\rm eff}$ are greater for
the positron in higher partial waves.


\section{Summary and outlook}

In this paper we have formulated a many-body theory approach which
accounts for the main correlation effects in positron-atom interactions.
These are (i) polarisation of the target by the positron, (ii) virtual
positronium formation, and (iii) strong enhancement of the electron-positron
contact density due to their Coulomb interaction. The key development
for an accurate description of (ii) and (iii) is the
summation of the ladder diagram series and calculation of the
electron-positron vertex function. B-spline basis sets and extrapolation over
the orbital angular momenta are used to achieve convergence of the sums over
the electron and positron intermediate states. The method can be applied to a
range of problems such as positron scattering, annihilation and formation of
bound states.

Although our main interest is in exploring many-electron targets,
the method has been first tested for hydrogen, where accurate benchmark
data exist for the scattering phaseshifts and annihilation rates.
In the case of hydrogen the present formalism is exact. Numerically,
excellent agreement with accurate variational calculations for the phase
shifts has been obtained, together with a good agreement for the
annihilation parameter $Z_{\rm eff}$. The calculation of the most difficult
part of the correlation potential, which contains the vertex function,
for many-electron atoms is only marginally more difficult than for
hydrogen. Therefore, we expect that application of our many-body theory to
the problems of positron scattering and annihilation on noble-gas
atoms and binding to halogen ions \cite{Ludlow:04a,Ludlow:04b} should yield
accurate results. In particular, we would like to re-examine and improve the
accuracy of the many-body theory predictions \cite{Dzuba:95} of positron
binding energies to the $ns^2$ atoms such as Mg, Cd and Zn.

The advantage of many-body theory methods is their physical transparency.
It allows one to distinguish between different physical mechanisms and
compare their relative importance. Thus, we saw that virtual Ps formation 
in positron-hydrogen scattering is just as important as the target
polarisation. Correlation corrections to the annihilation vertex, which are
physically related to the virtual Ps formation, are even more important.
They enhance the annihilation rate in hydrogen by a factor of 5 or more,
depending on the positron partial wave. Such vertex corrections depend weakly
on the positron energy, and the enhancement they produce is practically
the same for various positron wavefunctions. Therefore, one could use
the average enhancement factors derived for isolated atoms to obtain reliable
annihilation rates for atoms placed in different environments,
provided that a sufficiently accurate single-particle positron wavefunction
is available. Of course, different atomic subshells will be
characterised by different enhancement factors. However, they can all be
determined in an many-body calculations of the type described
in this paper, and serve as an input for the calculations of
positron annihilation in molecules or condensed matter.

In this work we have analysed the spatial dependence of the nonlocal
correlation corrections to the annihilation vertex. We have also derived
the equivalent local enhancement factor and compared it with an expression
used in condensed matter calculations. Similar comparisons for
larger many-electron targets may test various forms of enhancement factors
in a much greater range of electron densities.

The rapid development of computers over the past few decades seems
to have favoured theoretical methods other than many-body theory.
Such methods often rely more on the computer power and numerical
techniques than on the physical insight. They often appear to be
``more exact'' than the sophisticated but explicitly approximate
many-body theory approaches, and promise improved results due to shear
growth of computer power.
Their drawback is that they do not always increase one's physical
understanding of the problem. We believe that a further theoretical
development of many-body methods combined with a judicial use
of computers is a healthy alternative.

\acknowledgments
GG is grateful to V. V. Flambaum who drew the positron-atom problem to his
attention while at the University of New South Wales (Sydney),
and to W. R. Johnson for pointing out the advantages of B-spline
bases. The work of JL has been supported by the award from the
Department of Employment and Learning (Northern Ireland). We also thank
P. Van Reeth for providing data for positron annihilation on hydrogen
in numerical form.

\appendix

\section{Technical details}

For a positron interacting with a spherically-symmetric target,
both the self energy $\Sigma _E ({\bf r},{\bf r}')$ and the correlation
correction to the annihilation vertex $\Delta _E ({\bf r},{\bf r}')$
can be expanded in partial waves, e.g.,
\begin{equation}\label{eq:part}
\Sigma _E ({\bf r},{\bf r}')=\frac{1}{rr'}\sum _{\lambda =0}^{\infty}\sum 
_{\mu =-\lambda}^{\mu =\lambda }Y_{\lambda\mu}(\Omega )
\Sigma _E ^{(\lambda )}(r,r^{\prime})
Y_{\lambda\mu}^*(\Omega ').
\end{equation}
Defining the volume element $d{\bf r}=r^2drd\Omega $, and the
positron wavefunction with orbital angular momentum $l$,
$\varphi _{\eps }({\bf r})=r^{-1}P_{\eps l}(r) Y_{lm}(\Omega)$, we obtain
the matrix element $\langle\eps |\Sigma |\eps '\rangle$ as,
\begin{eqnarray}\label{eq:matrel}
\langle\eps |\Sigma _E|\eps '\rangle &=&
\int \varphi _\eps ^*({\bf r})\Sigma _E({\bf r},{\bf r}')
\varphi _{\eps '}({\bf r}')d{\bf r} \nonumber \\
&=&\int P_{\eps l}(r)\Sigma _E ^{(l)}(r,r')
P_{\eps ' l}(r')drdr'.
\end{eqnarray}
The angular reduction of the various diagrams in
$\langle \eps |\Sigma _E|\eps '\rangle $ and
$\langle \eps |\Delta _E|\eps \rangle $ is simplified by the use of graphical
techniques for performing angular momentum algebra \cite{vars}. 
The final expressions for the diagrams in terms of reduced matrix
elements, $3j$ and $6j$ symbols are given below. 

The reduced Coulomb matrix element is defined as,
\begin{eqnarray}\label{eq:Vl}
\langle 3,4\|V_l\|2,1 \rangle &=&\sqrt{[l_1][l_2][l_3][l_4]}
\left({l_1\atop 0}{l\atop 0}{l_3\atop 0}\right)
\left({l_2\atop 0}{l\atop 0}{l_4\atop 0}\right)\nonumber\\
&\times&\int P_{\eps_3l_3}(r_1)P_{\eps_4l_4}(r_2)
\frac{r_{<}^{l}}{r_{>}^{l+1}}P_{\eps_2l_2}(r_2)P_{\eps_1l_1}(r_1)
dr_1dr_2,
\end{eqnarray}
where $[l_1]\equiv 2l_1+1$, etc. Note that instead of including the minus
sign in the reduced Coulomb matrix element when it involves the
positron, we account for it in the overall sign factor for the diagram
(see below). The reduced Coulomb matrix element for an
electron-positron pair coupled into a total angular momentum $J$ is given by
\begin{eqnarray}\label{eq:VJ}
\langle 3,4 \|V^{(J)}\|2,1\rangle &=&\sum _l (-1)^{J+l}
\langle 3,4\|V_l\|2,1\rangle \left\{
{J\atop l}{l_3\atop l_2}{l_4\atop l_1}\right\} .
\end{eqnarray}
This expression is similar to the `exchange' matrix
element that one meets in all-electron problems \cite{Amusia:75}.

The sum of the ladder diagram series (the vertex function) is calculated
via the matrix equation (\ref{eq:gamma}), which must be solved for all possible
total angular momenta $J$ of the electron-positron pair:
\begin{equation}\label{eq:GamJ}
\langle \nu _2,\mu _2\|\Gamma _E^{(J)}\|\mu _1,\nu _1 \rangle=-
\langle \nu _2,\mu _2\|V^{(J)}\|\mu _1,\nu _1\rangle -
\sum_{\nu,\,\mu }\frac{\langle\nu _2,\mu _2 \|V^{(J)}\|\mu ,\nu \rangle
\langle \nu ,\mu \|\Gamma _E^{(J)}\|\mu _1,\nu _1 \rangle}
{E-\eps_\nu-\eps _\mu }.
\end{equation}
In the Appendix, the state labels $\nu $, $\nu _1$, etc., refer to the
positron orbitals $\eps _\nu l_\nu $, $\eps _{\nu _1}l_{\nu _1}$, etc.
Similarly, $\mu $, $\mu _1$, etc., label excited-state electron orbitals.
Electron orbitals occupied in the target ground state (holes) are labeled by
Latin indices ($n$). When B-spline basis states are used,
Eq. (\ref{eq:GamJ}) is a finite-dimension matrix equation solved by matrix
inversion.

The self-energy diagrams can be expressed in terms of these matrix elements.
For closed-shell atoms, each loop in the diagram gives a spin factor of 2.
This factor should be omitted for hydrogen which has only one electron in
the $1s$ orbital. The sign factor for each diagram is $(-1)^{a+b+c}$,
where $a$ is the number of hole lines, $b$ is the number of electron-hole loops
and $c$ is the number of positron-electron Coulomb interactions.
In the expressions below we label the incident positron angular momentum by
$l_p$. This is also the orbital angular momentum of the external lines
$\eps $, $\eps '$ of the diagrams.

The 2nd-order self-energy diagram, Fig. \ref{fig:sig}(a), is given by
\begin{equation}\label{eq:Sig2r}
2\sum_{\nu ,\,\mu , \,n}\sum _l
\frac{\langle \eps ',n\|V_l\|\mu , \nu\rangle
\langle \nu,\mu\|V_l \|n,\eps \rangle}
{[l][l_p](E+ \eps_n -\eps_\nu -\eps_ \mu )}.
\end{equation}
The virtual-Ps contribution to $\Sigma _E$, Fig. \ref{fig:Ps},
is obtained after finding the vertex function as follows:
\begin{equation}\label{eq:SigGr}
2\sum_{\nu_i,\,\mu_i,\,n}\sum _J
\frac{[J]\langle \eps ',n \|V^{(J)}\|\mu_2,\nu_2\rangle
\langle \nu_2,\mu_2\|\Gamma _{E+\eps _n}^{(J)}\|\mu_1, \nu_1\rangle
\langle \nu_1,\mu_1\|V^{(J)}\|n,\eps \rangle}
{[l_p](E+\eps_n-\eps_{\nu_1}-\eps_{\mu_1})
(E+\eps_n-\eps_{\nu_2}-\eps_{\mu_2})}.
\end{equation}

Matrix elements of the annihilation $\delta $-function are defined similarly
to the Coulomb ones, using the expansion of the $\delta ({\bf r}_1-{\bf r}_2)$
in terms of spherical harmonics \cite{vars}. The reduced matrix element then
is
\begin{eqnarray}\label{eq:dl}
\langle 3,4\|\delta_l \|2,1\rangle &=&\frac{[l]}{4\pi }
\sqrt{[l_1][l_2][l_3][l_4]}
\left({l_1\atop 0}{l\atop 0}{l_3\atop 0}\right)
\left({l_2\atop 0}{l\atop 0}{l_4\atop 0}\right)\nonumber \\
&\times &\int P_{\eps_3l_3}(r)P_{\eps_4l_4}(r)P_{\eps_2l_2}(r)P_{\eps_1l_1}(r)
r^{-2}dr,
\end{eqnarray}
and the matrix element for an electron-positron pair coupled into a total
angular momentum $J$ is given by,
\begin{eqnarray}\label{eq:dJ}
\langle 3,4\|\delta ^{(J)}\|2,1 \rangle =\sum _l (-1)^{J+l}
\langle 3,4\|\delta _l\|2,1\rangle \left\{
{J\atop l}{l_3\atop l_2}{l_4\atop l_1}\right\} .
\end{eqnarray}

The various diagrams contributing to $Z_{\rm eff}$ can then be expressed in
terms of these matrix elements. As with the self energy diagrams, the
factor of 2 which results from closed loops, must be removed
for hydrogen. Note that in the diagrams below, we also include factors of 2
to account for the mirror images of those diagrams that are not symmetric.
To produce correct values of $Z_{\rm eff}$ the expressions given below must
be multiplied by the positron normalisation factor (\ref{eq:fact}).

The 0th-order diagram, Fig. \ref{fig:Zeff}(a), is a sum
of simple radial integrals over all hole orbitals $n$,
\begin{equation}\label{eq:Zeff0r}
2\sum_n\frac{[l_n]}{4\pi}\int P_{\eps l_p}^2(r)P_{\eps _nl_n}^2(r)r^{-2}dr.
\end{equation}
The 1st-order contribution, Fig. \ref{fig:Zeff}(b) and (c), is given by
\begin{equation}\label{eq:Zeff1r}
-4\sum_{\nu,\,\mu,\,n}\sum_l
\frac{\langle \eps ,n\|\delta_l\|\mu,\nu\rangle
\langle \nu,\mu\|V_l \|n,\eps\rangle}
{[l][l_p](E+\eps_n-\eps _\nu -\eps _\mu)}.
\end{equation}
Expressions for the remaining four contributions, Fig. \ref{fig:Zeff1}(c)--(f),
are:
\begin{equation}\label{eq:Zeff2r}
2\sum_{\nu _i,\,\mu_i,\,n}\sum_J
\frac{[J]\langle \eps ,n \|V^{(J)}\|\mu_2,\nu_2\rangle
\langle \nu_2,\mu_2\|\delta^{(J)}\|\mu_1,\nu_1\rangle
\langle \nu_1,\mu_1\|V^{(J)}\|n,\eps\rangle}
{[l_p](E+\eps_n-\eps_{\nu_2}-\eps_{\mu_2})
(E+\eps_n-\eps_{\nu_1}-\eps_{\mu_1})},
\end{equation}
\begin{equation}\label{eq:ZeffG}
-4\sum_{\nu_i,\,\mu_i,\,n}\sum_J
\frac{[J]\langle \eps ,n \|\delta^{(J)}\|\mu_2,\nu_2\rangle
\langle \nu_2,\mu_2\|A _{E+\eps _n}^{(J)}\|n,\eps\rangle}
{[l_p](E+\eps_n-\eps_{\nu_2}-\eps_{\mu_2})},
\end{equation}
\begin{equation}\label{eq:ZeffGV}
4\sum_{\nu_i,\,\mu_i,\,n}\sum_J
\frac{[J]\langle \eps ,n \|V^{(J)}\|\mu _3,\nu_3\rangle
\langle \nu_3,\mu_3\|\delta^{(J)}\|\mu_2,\nu_2\rangle
\langle \nu_2,\mu_2\|A_{E+\eps _n}^{(J)}\|n,\eps \rangle }
{[l_p](E+\eps_n-\eps_{\nu_3}-\eps_{\mu_3})
(E+\eps_n-\eps_{\nu_2}-\eps_{\mu_2})},
\end{equation}
\begin{equation}\label{eq:ZeffGG}
2\sum_{\nu_i,\,\mu_i,\,n}\sum_J
\frac{[J]\langle \eps ,n \|A_{E+\eps _n}^{(J)}\|\mu _3,\nu_3\rangle
\langle \nu_3,\mu_3\|\delta^{(J)}\|\mu_2,\nu_2\rangle
\langle \nu_2,\mu_2\|A_{E+\eps _n}^{(J)}\|n,\eps \rangle }
{[l_p](E+\eps_n-\eps_{\nu_3}-\eps_{\mu_3})
(E+\eps_n-\eps_{\nu_2}-\eps_{\mu_2})},
\end{equation}
where we have introduced
\begin{equation}\label{eq:A}
\langle \nu_2,\mu_2\|A_{E+\eps _n}^{(J)}\|n,\eps\rangle =
\sum_{\nu_1,\,\mu_1}
\frac{\langle \nu_2,\mu_2\|\Gamma _{E+\eps _n}^{(J)}\|\mu_1,\nu_1\rangle
\langle \nu_1,\mu_1\|V^{(J)}\|n,\eps\rangle}
{E+\eps_n-\eps_{\nu_1}-\eps_{\mu_1}}
\end{equation}
This amplitude not only helps to make the expressions above
more compact. It also provides a practical way of computing these diagrams.
Thus, without a preliminary evaluation of
$ \langle \nu_2,\mu_2\|A_{E+\eps _n}^{(J)}\|n,\eps\rangle $, diagram (f)
in Fig. \ref{fig:Zeff1} contains eight-fold summation over hundreds
of electron and positron basis states.



\begin{thebibliography}{10}
\bibitem{correction}This is true to the extent that we neglect the small
difference between the true and HF electron densities, and 
that between $|\psi _\eps ({\bf r})|^2$ and $\rho_p({\bf r})$.
\end{thebibliography}
\end{document}